\documentclass[twocolumn, times]{aastex631}

\usepackage{amsmath}
\usepackage{xspace}
\usepackage{comment}

\usepackage{autobreak}

\shorttitle{Kinematics of streamers in IRAS~16544$-$1604}
\shortauthors{Kido et al.}

\begin{document}
\title{Early Planet Formation in Embedded Disks (eDisk) XXI: Limited role of streamers in mass supply to the disk in the Class 0 protostar IRAS~16544-1604}

\correspondingauthor{Miyu Kido}
\email{k3394334@kadai.jp}

\author[0000-0002-2902-4239]{Miyu Kido}
\affiliation{Academia Sinica Institute of Astronomy \& Astrophysics (ASIAA), 11F of Astronomy-Mathematics Building, AS/NTU, No.1, Sec. 4, Roosevelt Rd, Taipei 106319, Taiwan}
\affiliation{Department of Physics and Astronomy, Graduate School of Science and Engineering, Kagoshima University, 1-21-35 Korimoto, Kagoshima,Kagoshima 890-0065, Japan}

\author[0000-0003-1412-893X]{Hsi-Wei Yen}
\affiliation{Academia Sinica Institute of Astronomy \& Astrophysics (ASIAA), 11F of Astronomy-Mathematics Building, AS/NTU, No.1, Sec. 4, Roosevelt Rd, Taipei 106319, Taiwan}

\author[0000-0003-4361-5577]{Jinshi Sai} 
\affiliation{Academia Sinica Institute of Astronomy \& Astrophysics (ASIAA), 11F of Astronomy-Mathematics Building, AS/NTU, No.1, Sec. 4, Roosevelt Rd, Taipei 106319, Taiwan}

\author[0000-0003-0845-128X]{Shigehisa Takakuwa}
\affiliation{Department of Physics and Astronomy, Graduate School of Science and Engineering, Kagoshima University, 1-21-35 Korimoto, Kagoshima,Kagoshima 890-0065, Japan}
\affiliation{Academia Sinica Institute of Astronomy \& Astrophysics (ASIAA), 11F of Astronomy-Mathematics Building, AS/NTU, No.1, Sec. 4, Roosevelt Rd, Taipei 106319, Taiwan}

%================================================
\author[0000-0003-0998-5064]{Nagayoshi Ohashi}
\affiliation{Academia Sinica Institute of Astronomy \& Astrophysics (ASIAA), 11F of Astronomy-Mathematics Building, AS/NTU, No.1, Sec. 4, Roosevelt Rd, Taipei 106319, Taiwan}

\author[0000-0003-3283-6884]{Yuri Aikawa}
\affiliation{Department of Astronomy, Graduate School of Science, The University of Tokyo, 7-3-1 Hongo, Bunkyo-ku, Tokyo 113-0033, Japan}

\author[0000-0002-8238-7709]{Yusuke Aso}
\affiliation{Korea Astronomy and Space Science Institute, 776 Daedeok-daero, Yuseong-gu, Daejeon 34055, Republic of Korea}
\affiliation{Department of Astronomy and Space Science, University of Science and Technology, 217 Gajeong-ro, Yuseong-gu, Daejeon 34113, Republic of Korea}

\author[0000-0002-8591-472X]{Christian Flores}
\affiliation{Academia Sinica Institute of Astronomy \& Astrophysics (ASIAA), 11F of Astronomy-Mathematics Building, AS/NTU, No.1, Sec. 4, Roosevelt Rd, Taipei 106319, Taiwan}

\author[0000-0002-9143-1433]{Ilseung Han}
\affiliation{Department of Astronomy and Space Science, University of Science and Technology, 217 Gajeong-ro, Yuseong-gu, Daejeon 34113, Republic of Korea}
\affiliation{Korea Astronomy and Space Science Institute, 776 Daedeok-daero, Yuseong-gu, Daejeon 34055, Republic of Korea}
\affiliation{Department of Earth Science Education, Seoul National University, 1 Gwanak-ro, Gwanak-gu, Seoul 08826, Republic of Korea}
\affiliation{Institut de Ci\`encies de l'Espai (ICE-CSIC), Campus UAB, Can Magrans S/N, E-08193 Cerdanyola del Vall\`es, Catalonia, Spain}

\author[0000-0003-2777-5861]{Patrick M. Koch}
\affiliation{Academia Sinica Institute of Astronomy \& Astrophysics (ASIAA), 11F of Astronomy-Mathematics Building, AS/NTU, No.1, Sec. 4, Roosevelt Rd, Taipei 106319, Taiwan}

\author[0000-0003-4022-4132]{Woojin Kwon}
\affiliation{Department of Earth Science Education, Seoul National University, 1 Gwanak-ro, Gwanak-gu, Seoul 08826, Republic of Korea}
\affiliation{SNU Astronomy Research Center, Seoul National University, 1 Gwanak-ro, Gwanak-gu, Seoul 08826, Republic of Korea}
\affiliation{The Center for Educational Research, Seoul National University, 1 Gwanak-ro, Gwanak-gu, Seoul 08826, Republic of Korea}

\author[0000-0003-3119-2087]{Jeong-Eun Lee}
\affiliation{Department of Physics and Astronomy, Seoul National University, 1 Gwanak-ro, Gwanak-gu, Seoul 08826, Republic of Korea}

\author[0000-0002-7402-6487]{Zhi-Yun Li}
\affiliation{Astronomy Department and Virginia Institute for Theoretical Astronomy, University of Virginia, 530 McCormick Rd., Charlottesville, Virginia 22904, USA}

\author[0000-0002-4540-6587]{Leslie W. Looney}
\affiliation{Department of Astronomy, University of Illinois, 1002 West Green St, Urbana, IL 61801, USA}

\author[0000-0002-0554-1151]{Mayank Narang}
\affiliation{Academia Sinica Institute of Astronomy \& Astrophysics (ASIAA), 11F of Astronomy-Mathematics Building, AS/NTU, No.1, Sec. 4, Roosevelt Rd, Taipei 106319, Taiwan}

\author[0000-0003-1549-6435]{Kazuya Saigo}
\affiliation{Department of Physics and Astronomy, Graduate School of Science and Engineering, Kagoshima University, 1-21-35 Korimoto, Kagoshima,Kagoshima 890-0065, Japan}

\author[0000-0002-0549-544X]{Rajeeb Sharma}, 
\affiliation{Niels Bohr Institute, University of Copenhagen, \O ster Voldgade 5-7, 1350, Copenhagen, Denmark}

\author[0000-0003-0334-1583]{Travis J. Thieme}
\affiliation{Academia Sinica Institute of Astronomy \& Astrophysics (ASIAA), 11F of Astronomy-Mathematics Building, AS/NTU, No.1, Sec. 4, Roosevelt Rd, Taipei 106319, Taiwan}

\author[0000-0001-8105-8113]{Kengo Tomida}
\affiliation{Astronomical Institute, Graduate School of Science, Tohoku University, Sendai 980-8578, Japan}

\author[0000-0001-5058-695X]{Jonathan P. Williams}
\affiliation{Institute for Astronomy, University of Hawai‘i at Mānoa, 2680 Woodlawn Dr., Honolulu, HI 96822, USA}

\begin{abstract}
%We present detailed studies of the physical properties of streamers in the Class 0 protostar IRAS 16544-1604 and the environment of its parental dense core with combined ALMA and JCMT data.  
Asymmetric and narrow infalling structures, often called streamers, have been observed in several Class 0/I protostars, which is not expected in the classical star formation picture. Their origin and impact on the disk formation remain observationally unclear. 
By combining data from the James Cleark Maxwell Telescope (JCMT) and Atacama Large Millimeter/submillimeter Array (ALMA), we investigate the physical properties of the streamers and parental dense core in the Class 0 protostar, IRAS~16544$-$1604.
Three prominent streamers associated to the disk with lengths between 2800 to 5800 au, are identified on the northern side of the protostar in the C$^{18}$O emission. Their mass and mass infalling rates are estimated to be in the range of (1--4)$\times$10$^{-3}$~$M_\odot$ and (1--5)$\times$10$^{-8}$~$M_\odot$~yr$^{-1}$, respectively. Infall signatures are also observed in the more diffuse extended protostellar envelope observed with the ALMA from the comparison to the infalling and rotating envelope model.
The parental dense core detected by the JCMT observation has a mass of $\sim$0.5~$M_\odot$, sub to transonic turbulence of $\mathcal{M}$ $=$ 0.8--1.1, and a mass-to-flux ratio of 2--6.
Our results show that the streamers in IRAS~16544-1604 only possess 2\% of the entire dense core mass and contribute less than 10\% of the mass infalling rate 
%of (1--5)$\times$10$^{-6}$ $M_\odot$ yr$^{-1}$ 
of the protostellar envelope. Therefore, the streamers in IRAS~16544-1604 play a minor role in the mass accretion process onto the disk, in contrast to those streamers observed in other sources and those formed in numerical simulations of collapsing dense cores with similar turbulence and magnetic field strengths.  

\end{abstract}

\section{Introduction} \label{sec:intro}
Protostars are formed when dense cores collapse under their self-gravity \citep{1969Larson}. Simultaneously, protostellar disks composed of gas and dust are formed around these forming protostars  \citep[]{1977Shu, 1984Terebey}.
In this early stage of star formation (Class 0 and I), the remaining gas in the dense core exists as an envelope, which actively accretes onto the disks.
%In the Class 0 stage, disk substructures indicative of the presence of planets, such as rings and gaps have not been observed; instead, filled disks are detected \citep[e.g.,][]{2023Ohashi}. 
%\citep[e.g.,][N. T. Phuong et al.; Han et al. in preparation]{2023van'tHoff, 2023Aso, 2023Thieme, 2023Sharma, 2023Sai, 2023Lin, 2023Ohashi, 2023Narayanan, 2024Gavino, 2024Santamaria-Miranda, 2024Encalada}. 
As the protostellar system evolves from the Class I to Class II stages, protoplanets may form rapidly within the disks, as suggested by the limited substructures in young protostellar disks \citep[e.g.,][]{2023Ohashi, 2023Han} compared to ubiquitous ring-gap structures observed in Class II disks \citep[e.g.,][]{2018Andrews}. 
%\citep[e.g.,][]{2017Sheehan, 2018Sheehan, 2020Sheehan, 2020Segura-Cox, 2023Yamato, 2023Flores, 2024Shoshi}. 
The kinematics and chemical properties of Class 0/I embedded disks are strongly influenced by mass accretion from the envelope \citep{2024Tobin}.
Thus, investigating the accretion process onto the disks in the young protostellar stage is essential to understand the planet formation process in the disks.
%The ALMA Large program ``Early Planet Formation in Embedded Disks (eDisk, \cite{2023Ohashi})'' conducted high-resolution radio observations toward Class 0/I protostars---
%High-resolution radio observations toward Class 0/I protostars have been conducted, such as ALMA Large program ``Early Planet Formation in Embedded Disks (eDisk, \cite{2023Ohashi})''. %In the eDisk program, 19 young embedded sources are observed at an angular resolution of 0\farcs04 in the 1.3-mm band to explore the presence of substructures in the early phase of the star formation. 
%From their observations, young objects with low bolometric temperatures were found to have filled dust disk \citep{2023Kido, 2023Lin, }, whereas the two objects with the highest bolometric temperatures showed ring+gap structures in dust continuum emission \citep{2023Yamato, 2023Flores}. This suggests that planet formation may already be initiated at the protostellar stage, indicating the necessity of focusing on not only the structures in the protoplanetary disk of the Class II stage but also the structures in the protostellar disk including how the gas accretes onto the disk in the active accretion phase to understand the comprehensive planet formation process.

Recent high-resolution (sub-)millimeter radio observations detected asymmetric, filament-like structures connected to protostellar disks around several protostars at various evolutionary stages from Class 0 to II, even though the conventional pictures of star formation assume spherically symmetric collapsing gas around a protostellar system  
\citep[e.g.,][and \cite{2023Pineda} for a recent review]{2012Tobin, 2014Tokuda, 2020Pineda, 2020Sai, 2021Huang, 2023Flores, 2023Aso, 2023Gupta, 2024Lee}.
%\citep[e.g.,][]{2012Tobin, 2012Tang, 2014Tokuda, 2018Tokuda, 2019Yen, 2019Akiyama, 2019Alves, 2020Pineda, 2020Alves, 2023Aso, 2021Ginski, 2020Huang_a, 2020Sai, 2021Huang, 2022Huang, 2023Huang, 2021Garufi, 2022Garufi, 2022Valdivia, 2023Valdivia, 2024Valdivia, 2023Lee, 2023Yamato, 2023Aso, 2023Harada, 2023Gupta, 2023Mercimek, 2024Zurlo, 2024Hales}. 
These structures are often called ``streamers'', characterized by narrow gas flows with velocity structures suggestive of infalling motions, and they are presumed to supply both mass and angular momentum to the disks. %\cite[e.g.,][for a recent review]{2023Pineda}. 
%They are detected in various lengths and molecule lines and thus streamers have different kinetic and chemical properties for each sources. 
%in addition to the isotropic infall in the classical star formation model (REF).  
%In many papers, it is said 
It has been suggested that streamers can play an important role in providing mass to the disk \citep[e.g.,][]{2022Valdivia, 2022Thieme}. 
%because the mass infall rates are similar to that of global envelope component and/or the infall rate is larger than the accretion rate from disk to star.
Similar streamers are often seen in numerical simulations of the gravitational collapse of magnetized and turbulent dense cores under various physical conditions. 
%They are often formed under various physical conditions,
%, and supply the mass to the disks 
%regardless of the strength of the turbulence, the mass of the dense core, and the presence of the global rotation
These simulations highlight the critical role of streamers in delivering mass to the disk and their significance to the disk formation process \citep{2015Seifried, 2017Kuffmeier, 2019Lam, 2023Tu, 2024Lebreuilly}, where the mass delivered by streamers can reach 0.5 $M_\odot$ in some cases \citep[e.g.,][]{2023Kuffmeier}.
However, observationally, the formation mechanism of streamers and their impact on the disk formation remain unclear.  

%Their detections suggests that, in addition to the isotropic accretion component proposed in the classical model (REF), there are also anisotoropic accretion components. So far, they have been detected in low-mass star formation at various evolutionary stages, Class 0 (REF), Class I (REF), and Class II (REF) as well as high-mass star formation (REF).  
%\cite{2024Valdivia-Mena} have found that the fraction of the streamer candidates in YSOs with the NOEMA interferometer and the IRAM 30m single dish, and  the resulting fraction is $\sim$40 \%. On the other hand, although the number of detected streamers is still small, they have been observed in various regions and evolutionary stages, indicating the need to understand their role in the star formation process. The detection of streamers gradually increases, and streamers are the key to understand comprehensive star and planet formation process.
%Hence, it leads to the questions: how important are streamers in disk formation? and what is the origin of streamers? 
%In numerical simulations of the gravitational collapse, streamer-like components are generated and supply the mass to the disk (REF). 2015Seifried reported that these structures are found in the magnetized dense core regardless of the strength of the turbulence, the mass of the dense core, and the presence of the global rotation. In addition, 2023Tu are also reported as main contributor of the disk formation.

IRAS~16544$-$1604 (hereafter IRAS~16544) is a Class 0 protostar located in the Ophiuchus North region \cite[$d$ $=$ 151~pc;][]{2020Zucker}. The bolometric temperature is estimated to be 50~K \citep{2023Ohashi}. It is deeply embedded in a dense core elongated along the north-south direction with a size of 20,000~au \citep{2000Vallee, 2014Bertrang, 2024Yen}, and is associated with a bipolar outflow along the northwest-southeast direction \citep[]{1996Wu, 2003Vallee, 2007Vallee, 2014Dunham}. 
IRAS~16544 is one of the targets of the Atacama Large Millimeter/submillimeter Array (ALMA) large programs ``Fifty AU Study of the chemistry in the disk/envelope system of Solar-like protostars (FAUST; \cite{2022Imai})''. They revealed that IRAS~16544 harbors a hybrid chemistry with complex organic molecules detected around the protostar and warm carbon-chain chemistry molecules distributed on the envelope scale of 1000~au.
%, which means that complex organic molecules are detected around the protostar while warm carbon-chain chemistry molecules are extended on the envelope scale of 1000 au. 
The ALMA large programs ``Early Planet Formation in Embedded Disks (eDisk)'' also targeted IRAS~16544 \citep{2023Ohashi, 2023Kido}, and their observations showed a $\sim$30~au dusty disk oriented in the northeast-southwest direction and inclined by $\lesssim$73$\degr$ from the plane-of-sky.
%The ALMA observations revealed a $\sim$30 au dusty disk and abundant complex organic molecules around the protostar. The disk has a major axis along the northeast-southwest direction with a position angle of 45$\degr$ and is inclined by 73$\degr$ on the plane-of-sky with the near side in the northwest. 
The Keplerian disk has a radius of $\sim$55~au, and the protostellar mass is estimated to be 0.14~$M_\odot$ from the Keplerian rotation of the gas disk \citep{2023Kido}. 
Outside the Keplerian disk, several asymmetric, elongated components have been observed in C$^{18}$O ($J$ $=$ 2--1) emission in IRAS~16544.
The spatial and velocity structures of one of them (northeastern component) has been analyzed, and is consistent with being a streamer \citep{2023Kido}. 
This previous analyses did not, however, perform complete census of all the streamer candidates nor estimate their mass and mass infalling rates including the proper estimate of the missing fluxes. Thus, quantitative estimate of the importance of those streamers on the mass accretion to the star plus disk system has not yet been conducted.
%However, the importance of mass accretion process onto the protostellar disk in IRAS~16544 has not yet been investigated in detail.
%Its spatial and velocity structures can be described with the envelope model proposed by \cite{1976Ulrich, 1981Cassen} (hearafter CMU model). In this model, the material accretes onto the disk with the gravity of protostellar mass while conserving angular momentum.
%The systemic velocity is estimated to be 5.0 km s$^{-1}$ \citep{2022Imai, 2023Kido}.

To investigate the role of streamers in the star and disk formation process, we conducted follow-up observations using the James Clerk Maxwell Telescope (JCMT) toward IRAS~16544. In this paper, we utilize the unique combination of the JCMT, FAUST, and eDisk data that have overlapping {\it uv} coverages and result in a large spatial dynamical range, and we study the physical properties of the parental dense core and streamers in IRAS~16544 in detail. 
%In this paper, we study the physical properties of the parental dense core and streamers in IRAS~16544. By combining the data from the James Clerk Maxwell Telescope (JCMT) and ALMA, we investigate the role of the streamers in the star and disk formation process in the target source.
In Section~\ref{sec:obs}, we describe details of the JCMT and ALMA observations and data reduction. 
Our analysis to estimate physical parameters of the dense core and streamers is described in Section~\ref{sec:results}. 
Through these results, we discuss the origin of the streamers and their impact on the protostellar disk in Section~\ref{sec:discussion}.
Finally, we summarize our findings in Section~\ref{sec:summary}.
Comparison of the properties of the streamers probed with ALMA among all eDisk targets will be introduced in the forthcoming paper. 

\section{Observations and data reduction} \label{sec:obs}
\subsection{JCMT}
We observed IRAS~16544 in C$^{18}$O $J$ $=$ 2--1 (219.5603541 GHz) with the heterodyne receiver `\=U`\=u on JCMT in five nights between 2023 March and September as part of the eDisk follow-up project (Project codes: M23AP020, M23BP016; PI: J.~Sai). 
%Overall results are summarized in Sai et al. in preparation.
The sky opacity at 225 GHz ranged from 0.13 to 0.29. We mapped a square region of 120$\arcsec$$\times$120$\arcsec$ around the protostar with the raster scan mode. The bandwidth and spectral resolution were set to 250~MHz and 30.5~kHz (corresponding to 0.042~km~s$^{-1}$), respectively.
%(corresponding to $0.0418~\mathrm{km~s^{-1}}$), respectively. 
The full width half maximum (FWHM) beam size was 21$\arcsec$.
%at the rest frequency of the $\mathrm{C^{18}O}$ line. 
The obtained data were reduced with \texttt{Starlink} version 2023A. The antenna temperature was converted to the main beam temperature ($T_\mathrm{mb}$) with the main beam efficiency of 0.66. The image cubes were binned to a channel width of 91.6 kHz, corresponding to 0.125~km~s$^{-1}$. The rms noise level of 0.31 K per channel in $T_\mathrm{mb}$ was achieved.

%\subsection{ALMA}
%In this paper, three 12m array observations in Band 6, which were observed in ALMA Large program eDisk (project code: 2019.1.00261.L, PI: N. Ohashi), the Director's Discretionary Time (DDT) program (project code: 2019.A.00034.S, PI: J. J. Tobin), and another large program FAUST (project code: 2018.1.01205.L, PI: J. S. Yamamoto), were combined to increase the signal-to-noise ratio. 
\subsection{ALMA}
\label{datareduction:ALMA}
We utilize the ALMA 12-m and 7-m array Band 6 data obtained with the large program eDisk (project code: 2019.1.00261.L, PI: N.~Ohashi), the Director's Discretionary Time program (project code: 2019.A.00034.S, PI: J.~J.~Tobin), and the large program FAUST (project code: 2018.1.01205.L, PI: S.~Yamamoto). 
The details of the eDisk observations have been described in \cite{2023Ohashi} and \cite{2023Kido} while those of FAUST have been reported in \cite{2022Imai}.
%these observations have been described in \cite{2023Ohashi}, \cite{2023Kido} and \cite{2022Imai}.
%\subsection{eDisk}
%The baseline lengths range from 47 m to 12595 m and the details of the observational setup is described in \cite{2023Ohashi} and \cite{2023Kido}.
%\subsection{DDT}
%The baseline of 15 $\sim$ 1302 m was adopted for this observation, and see \cite{2023Ohashi} and \cite{2023Kido} for details of the observation.
%\subsection{FAUST}
%They observed IRAS 16544 at baseline lengths from 15 m to 1241 m . The observational details can be found in \cite{2022Imai}. 
%We combined all the data sets obtained with the 12-m and 7-m arrays from these programs.
%IRAS 16544 has been observed 6 times with 12-m array and 2 times with Atacama Compact Array (ACA): the 7-m array and the total power antennas (Leory+2021).
The combination of these data sets has the baseline lengths from 9 m to 12595 m and the maximum recoverable scale (MRS) of 6$\farcs$2. 

All the data were calibrated with the pipeline of the Common Astronomy Software Applications (CASA) package \citep{2022CASATeam}. We extracted the C$^{18}$O visibility data using the CASA task \texttt{mstransform} and subtracted the continuum using \texttt{uvcontsub}. 
%Self-calibration was not performed.
All the C$^{18}$O data were imaged together using the task \texttt{tclean} of CASA version 6.6 with the Briggs robust parameter of 2.0 and the mosaic gridder, resulting in a beam size of 0$\farcs$30$\times$0$\farcs$22 (P.A. $=$ 77$\degr$). The velocity resolution was set to 0.167~km~s$^{-1}$. The auto-masking function of clean was adopted with the sidelobe threshold of 1.5, noise threshold of 4.25, low noise threshold of 1.0, minimum beam fraction of 0.3, and negative threshold of 0.0. 
A primary beam correction was applied. 
The rms noise level of the final image cube was measured to be 1.35~mJy~beam$^{-1}$ ($=$ 0.52~K) within the radius of 10$\arcsec$ from the protostar in the line-free channels before the primary beam correction.

\subsection{JCMT and ALMA combined map}
%To understand the missing flux rate in streamers, we combined the ALMA 12-m+7-m and JCMT maps. 
We generated combined JCMT$+$ALMA 12-m and 7-m array maps of the C$^{18}$O emission.
The JCMT cube was first regridded with \texttt{imregrid} to have the same spatial and velocity axes as the ALMA cube and multiplied by the primary beam pattern of the ALMA map created in Section~\ref{datareduction:ALMA} with \texttt{immath}. 
%aligning the primary beam response of the JCMT map with that of the ALMA map. 
Then the ALMA and JCMT maps were combined using the \texttt{feather} task in CASA \citep[e.g.,][]{2017Cotton, 2023Plunkett}.
%We combined the ALMA and the JCMT maps without primary beam correction using \texttt{feather} task in CASA, and all
Finally, the combined map was divided by the ALMA primary beam pattern. The resulting beam size is 0$\farcs$30$\times$0$\farcs$22 (P.A. $=$ 77$\degr$). The rms noise level in the combined map was measured to be 1.46~mJy~beam$^{-1}$ ($=$ 0.56~K) within the radius of 10$\arcsec$ in the line-free channels before the primary beam correction.

\section{Analysis and Results} \label{sec:results}
We adopt three different maps for our analysis, i.e., the JCMT only, ALMA 12-m and 7-m combined (ALMA-only), and JCMT$+$ALMA 12-m and 7-m combined (feathered) maps (see Section~\ref{sec:obs}). %as described in Sec.~\ref{sec:obs}.
%The JCMT-only data is unaffected by missing flux, making it suitable for examining the physical environment of the dense core on scales of 10,000 au. Therefore, this data 
The JCMT-only map is used to study the mass and velocity field of the parental dense core (Section~\ref{subsec:dense core by JCMT}). 
The ALMA-only map resolves structures on scales of several hundred au and is used to investigate the spatial and velocity structure of the streamers (Sections~\ref{subsubsec:identified_streamers} and \ref{subsubsec:streamer properties}). 
%ALMA-only data is used to identify the streamers in Sec.~\ref{subsubsec:identified_streamers} and estimate their physical parameters in Sec.~\ref{subsubsec:streamer properties}.
The feathered map can recover the extended emission at high resolution. It is used to study the more diffuse extended protostellar envelope and assess the effect of missing flux in the ALMA-only map (Sections~\ref{subsubsec:missing flux} and \ref{subsec:model}).  
%of the streamers at the short UV spacing. We derived the emission without the effect of the missing flux. The Feather image is used to estimate the missing flux of streamers in Sec.~\ref{subsubsec:missing flux} and make Position-Velocity diagrams along major and minor axes in Sec.~\ref{subsec:model}.

\subsection{Dense core observed with JCMT}\label{subsec:dense core by JCMT}
\subsubsection{Morphology and Mass}\label{subsubsec:dense core mass}
\begin{figure*}[t]
\centering
\includegraphics[width=170mm, angle=0]{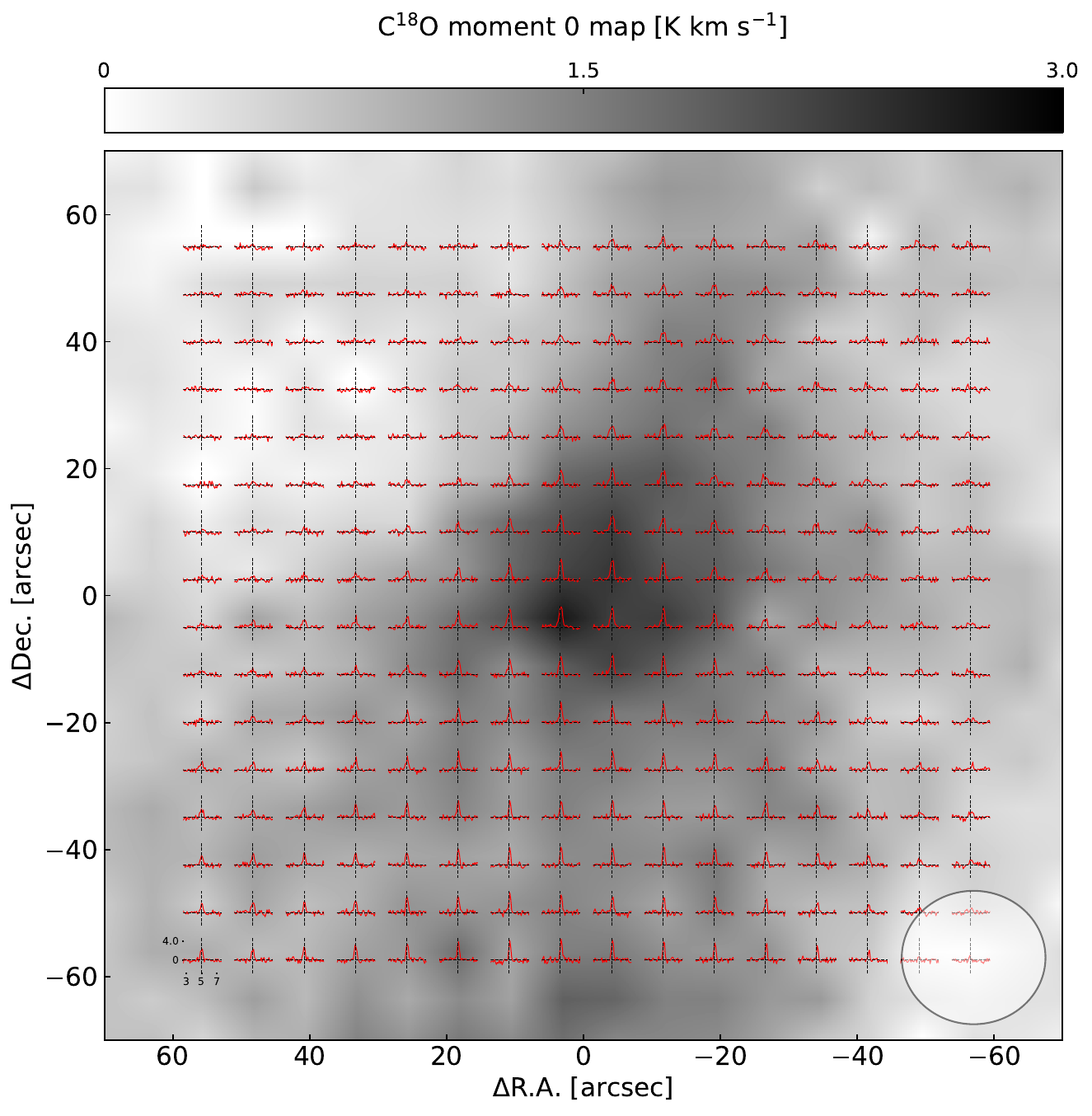}\\
\caption{Grid map of the C$^{18}$O ($J$ $=$ 2--1) spectra overlaid on its integrated intensity map (gray scale) of IRAS~16544 obtained with our JCMT observations (1$\sigma$ $=$ 0.16~K~km s$^{-1}$).
%The C$^{18}$O ($J$ = 2--1) spectral grid map for IRAS 16544.
%The background color scale image presents the integrated intensity map. 
In each grid, the vertical axis represents the main-beam brightness temperature in units of K, and the horizontal axis represents the LSR velocity in units of~km~s$^{-1}$ with the systemic velocity of 5~km~s$^{-1}$ indicated by a vertical dotted line.  The map center is set to be stellar position of ($\alpha_\mathrm{ICRS}$, $\delta_\mathrm{ICRS}$) $=$ (16$^\mathrm{h}$57$^\mathrm{m}$19$\fs$643, $-$16$^\mathrm{d}$09$^\mathrm{m}$24$\fs$016) as derived from the 2-dimensional Gaussian fitting to the 1.3-mm dust disk image \citep{2023Kido}.
A filled circle at the bottom-right corner shows the beam size of the JCMT observations of 21$\arcsec$.}
\label{fig_JCMTline}
\end{figure*}
\begin{comment}
\begin{figure*}[t]
\centering
\includegraphics[width=180mm, angle=0]{./figures/JCMT_mom0_columndensity_continuum.pdf}\\
\caption{Left: C$^{18}$O moment 0 map of the dense core in IRAS~16544, integrated from 4.0 to 6.0 km s$^{-1}$, obtained with the JCMT observations. Contour levels are 3, 5, 7, 9, 12 and 15$\sigma$, where 1$\sigma$ is 0.16 K km s$^{-1}$. Right: H$_2$ column density map (color) derived from the C$^{18}$O moment 0 map overlaid with the JCMT 850 $\mu$m continuum map (contour) from \cite{2024Yen}. Contour levels are 3, 10, 20, 30, 50, 70 and 90$\sigma$, where 1$\sigma$ is 4 mJy beam$^{-1}$.
%A white cross marks the position of IRAS 16544. 
Filled ellipses at the bottom-right corners show the beam sizes of the C$^{18}$O and dust continuum maps of 21$\arcsec$ and 14$\farcs$6, respectively.
The protostellar position is at zero offset in these maps.}
\label{fig_JCMTmom0}
\end{figure*}
\end{comment}
Figure~\ref{fig_JCMTline} displays the grid map of the C$^{18}$O (2--1) spectra taken by JCMT. All the spectra show a single velocity component, with the emission mainly detected within 1 km s$^{-1}$ with respect to the systemic velocity ($V_\mathrm{sys}$) of 5.0~km~s$^{-1}$. Since the peak brightness temperature of 4.8~K is much lower than the average dust temperature of 15~K of the dense core \citep{2013Launhardt}, the C$^{18}$O emission is most likely optically thin.

The background image of Figure \ref{fig_JCMTline} and Figure~\ref{fig_JCMT_nt}(a) show the integrated intensity (moment 0) map of the C$^{18}$O emission over the velocity range of $V_\mathrm{LSR}$ $=$ 4.0--6.0~km~s$^{-1}$ for IRAS~16544, obtained using JCMT.
The C$^{18}$O emission is elongated along the northwest-southeast direction. This direction is slightly shifted from that of the 850~$\mu$m dust continuum emission, which is elongated in the north-south direction \citep[black contours in Figure~\ref{fig_JCMT_nt}(a)]{2024Yen}, whereas the peak positions of the C$^{18}$O and dust continuum emission are coincident. 

We estimate the C$^{18}$O column density under the assumption of the local thermodynamic equilibrium (LTE) condition. 
%where $\nu$ is the frequency of C$^{18}$O of 219.560354 GHz, T$_{bg}$ is background radiation temperature of 2.725 K.
The excitation temperature is assumed to be 15 K, which is the average dust temperature in the envelope on a 60$\arcsec$ scale, as derived from Herschel observations by \cite{2013Launhardt}.
The mean molecular weight of 2.8 \citep{2008Kauffmann} and the C$^{18}$O abundance relative to H$_2$ of 1.7$\times$10$^{-7}$ \citep{1982Frerking} are adopted.  
%The C$^{18}$O abundance and the mean molecular weight ($\mu_{H_2}$) are assumed to be 1.7$\times$10$^{-7}$ \citep{1982Frerking} and 2.8, respectively.
%Einstein A coefficient (A$_{21}$) of 6.011$\times$10$^{-7}$ from LAMDA database \citep{Schoier2005}, the frequency of C$^{18}$O of 219.5603541 GHz, the background radiation temperature (T$_{bg}$) of 2.725 K.
%The right panel of Figure~\ref{fig_JCMTmom0} presents the estimated H$_2$ column density map. 
The mass within a box in the map of Figure~\ref{fig_JCMT_nt}(a) ($\sim$ 9000$\times$9000~au$^2$) is estimated to be $\sim$0.52 $M_{\odot}$. If the excitation temperature is assumed to be 10 or 30~K, the estimated core mass becomes 15\% higher. %0.60$M_{\odot}$. 
Our estimate is comparable to the previous estimates of 0.6--1.9~$M_{\odot}$ with radii of 1000-5000~au by \cite{2010Launhardt} and \cite{2014Bertrang} from the dust continuum emission. 

\subsubsection{Velocity Structures} \label{subsubsec:Linewidth}
\begin{figure*}[]
\centering
\includegraphics[width=180mm, angle=0]{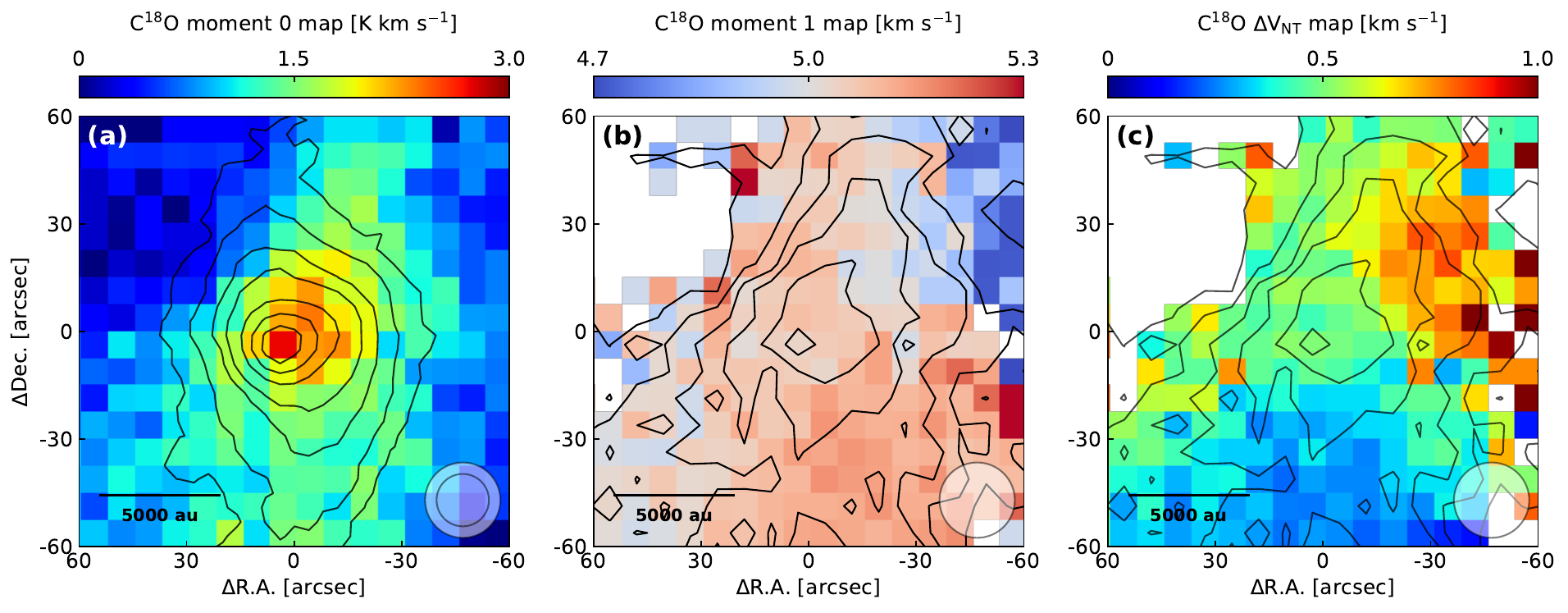}\\
\caption{
%Left: C$^{18}$O moment 1 map of the dense core in IRAS~16544, obtained with the JCMT observations. The emission higher than 3$\sigma$ is integrated to make this map. An open circle at the bottom-right corner shows the beam sizes of the C$^{18}$O observations of 21$\arcsec$.
%Right: Map of the FWHM non-thermal line width in the dense core in IRAS~16544 estimated from the C$^{18}$O line for the pixels with the peak C$^{18}$O intensity above 3$\sigma$. In the left and right panels, contours are the same as those in the left panel of Figure~\ref{fig_JCMTmom0}.
(a): C$^{18}$O moment 0 map of the dense core in IRAS~16544 obtained by the JCMT observations, integrated from 4.0 to 6.0~km~s$^{-1}$, overlaid with the JCMT 850~$\mu$m continuum map (contour) from \cite{2024Yen}. Contour levels are 3, 10, 20, 30, 50, 70 and 90$\sigma$, where 1$\sigma$ is 4~mJy~beam$^{-1}$.
(b): C$^{18}$O moment 1 map of the dense core (color) with overlaid the moment 0 map (contours).
The emission higher than 3$\sigma$ is integrated to make this map. 
Contours levels are 3, 5, 7, 9, 12 and 15$\sigma$, where 1$\sigma$ is 0.16~K~km s$^{-1}$ (left panel).
(c): Map of the FWHM non-thermal line width in the dense core estimated from the C$^{18}$O line for the pixels with the peak C$^{18}$O intensity above 3$\sigma$. Contours are the same as those in the center panel.
Filled circles at the bottom-right corners show the beam sizes of the C$^{18}$O and dust continuum maps of 21$\arcsec$ and 14$\farcs$6, respectively.
%The protostellar position is at zero offset in these maps.
}
\label{fig_JCMT_nt}
\end{figure*}
%Figure \ref{fig_JCMTline} displays the spectrum for each cell. 
%In the southern region, the line widths are narrow, and the peaks are sharp, while in the northen region, the line widths are broader, and the peaks are not as pronounced as in the south. 
%Overall, the spectra exhibit a single peak component at the systemic velocity of 5.0 km s$^{-1}$. 
%On the other hand, in the northwestern region of the dense core, double-peak profiles are  detected, with peaks in both the blue- and red-shifted velocities.
The Figure \ref{fig_JCMT_nt}(b) shows the mean velocity (moment 1) map of the C$^{18}$O emission, integrating the emission above 3$\sigma$. A weak velocity gradient on a 120$\arcsec$ scale can be seen from the northwest to southeast direction. In the central 60$\arcsec$ region, the velocity gradient is along the northeast-southwest direction. We measured the velocity gradient following the method in \cite{1993Goodman} and found that the resulting magnitude is $\sim$1.68$\pm$0.34~km~s$^{-1}$~pc$^{-1}$ with a position angle of 210$\pm$12$\degr$ within a radius of 60$\arcsec$. The direction of the core-scale velocity gradient is similar to that of the disk rotation with a position angle of 225$\degr$.

The C$^{18}$O line profiles in IRAS~16544 observed with JCMT are mostly symmetric with respect to the systemic velocity. Here, we divide the cloud into two regions, north and south, based on the declination offset of 0\arcsec. 
The broader line profiles are observed in the central and northern regions, while narrower profiles are seen in the southern region (Figure~\ref{fig_JCMTline}). %Most of the emission is moving with the systemic velocity of 5 km s$^{-1}$.
%, but slightly velocity gradient is observed, with the blue- and redshifted emissions located in the north and south regions.
We fit these spectra with a single Gaussian function. 
The fitting is performed for those spectra with a peak signal-to-noise ratio greater than 3. %to derive the Full-Width-Half-Maximum (FWHM) line width and centroid velocity. 
%Our Gaussian fitting to the spectrum at the central stellar position results in the central velocity and FWHM are 5.07 $\pm$0.01 km s$^{-1}$ and 0.53 km s$^{-1}$, respectively, approximately consistent with the systemic velocity estimated on $\sim$20$\arcsec$ scale with the ALMA data.
The centroid velocity and full width at half-maximum (FWHM) line width of the spectrum at the protostellar position are measured to be $V_\mathrm{LSR}$ $=$ 5.07$\pm$0.01~km~s$^{-1}$ and $\Delta V_\mathrm{total}$ $=$ 0.53$\pm$0.03~km~s$^{-1}$, respectively.
%The velocity gradients of the gas disk is detected along the northeast to the southwest and the southeast to the northwest, respectively. The direction is different from that of the disk because JCMT observed extended emission component, not disk scale.
%These difference can be seen in other sources \citep[e.g.,][]{2023Sai_gradient}. 
%Although the direction is different from that of the disk, the velocity gradient at the central stellar scale is consistent with the disk component.
We estimate the non-thermal gas motion by subtracting the thermal components from the measured C$^{18}$O linewidth as follows:
\begin{equation}
\Delta V_\mathrm{NT} = \sqrt[]{\Delta V_{\rm total}^2 - \frac{kT}{m_\mathrm{C^{18}O}}8\ln2},
\end{equation}
where $\Delta$$V_\mathrm{NT}$ is the FWHM of the non-thermal component,  $\Delta$$V_\mathrm{total}$ is the measured C$^{18}$O FWHM, $k$ is the Boltzmann
constant, $T$ is the kinetic temperature, and $m_\mathrm{C^{18}O}$ is the C$^{18}$O molecular mass \citep{2008Kauffmann}. $\Delta$$V_\mathrm{total}$ is from our Gaussian fitting to the spectra. 
We assume a uniform gas kinetic temperature of 15 K to derive the non-thermal gas motion, and a possible range of the kinetic temperature of 10--30~K to estimate the uncertainty.
%$T$ is assumed to be 10--30 K. %The results are shown in the left panel of Figure \ref{fig_JCMT_nt}.
%Setting the position of the central star at (0,0) and dividing the area into north and south (the gray dashed line in the left panel of Figure \ref{fig_JCMT_nt}) indicates the borderline between two region),
Figure \ref{fig_JCMT_nt}(c) shows the resultant FWHM map of the non-thermal motion in the dense core. 
It is larger in the northern region with an average $\Delta$$V_\mathrm{NT}$ of 0.61$^{+0.01}_{-0.02}$~km~s$^{-1}$, while it is lower in the southern region with an average $\Delta$$V_\mathrm{NT}$ of 0.41$^{+0.01}_{-0.03}$~km~s$^{-1}$. %The uncertainties here are from the range of the assumed temperature of 10--30 K. 
The Mach number, $\mathcal{M}$ $=$ $\sigma_\mathrm{NT}$/c$_s$ ($\Delta V_\mathrm{NT}$ $=$ $\sqrt[]{8\mathrm{log}2}\sigma_\mathrm{NT}$), is derived to be 1.1$^{+0.3}_{-0.2}$ in the northern region and 0.8$^{+0.2}_{-0.3}$ in the southern region. 
%, corresponding to transonic turbulence.
%, indicating that the existence of hydrodynamic waves within the core \citep{1975Arons, 2021Pineda}. 
%In contrast, the turbulence in the southern region is subsonic with a Mach number of 0.8$^{+0.2}_{-0.3}$. 
Thus, the dense core has subsonic to transonic non-thermal gas motions within a radius of 60$\arcsec$.

\subsubsection{B-field strength} \label{subsubsec:B-field}
The magnetic field structures in this dense core have been revealed with JCMT SCUBA-2/POL-2 observations \citep{2024Yen}. 
We adopt the Davis-Chandrasekhar-Fermi (DCF) method \citep{1951Davis,1953Chandrasekhar} to estimate the magnetic field strength in the dense core as follows:
\begin{equation}
|\boldsymbol{B}_\mathrm{pos}| = \xi\ \sqrt[]{4\pi\rho_\mathrm{gas}}\frac{\sigma_{\mathrm{NT}}}{\sigma_\theta},
\end{equation}
where $\xi$ is the correction factor to take account for variations of magnetic field and density structures along the line of sight, $\rho_\mathrm{gas}$ is the mean volume mass density, 
%(g cm-3; $\rho$ $=$ $\mu$$_{H_2}$n$_{H_2}$m$_H$, $\mu_{H_2}$$=$2.8), 
$\sigma_{\mathrm{NT}}$ is the mean 1$\sigma$ line width of non-thermal gas motion, and $\sigma_\theta$ is the angular dispersion of the magnetic fields. %{(radians)}, velocity dispersion (km/s? cm/s?)
$\sigma_{\mathrm{NT}}$ is calculated with the overlapping region of the 850~$\mu$m dust continuum emission and moment 0 map in Figure~\ref{fig_JCMT_nt}(a), and it is 0.18--0.20~km~s$^{-1}$.
$\sigma_\theta$ was estimated to be 7$\arcdeg$--13$\arcdeg$ within a radius of 60$\arcsec$ from the protostar with the JCMT POL-2 data in \cite{2024Yen}.
%and measured within a radius of 60$\arcsec$ from the protostar. It is enough to combine two data to calculate the magnetic field strength. 
$\xi$ of 0.5 is adopted since the angular dispersion is smaller than 25$\arcdeg$ \citep{2001Ostriker, 2001Heitsch}.
In the 850~$\mu$m dust continuum emission, the dense core is roundish with a diameter of 140$\arcsec$ (black contours in Figure~\ref{fig_JCMT_nt}(a)). Thus, we assume the depth of the dense core to also be 140$\arcsec$ and estimated the mean volume density to be 
%7.0(10K), 6.1(15K), 6.2(20K), 7.1(30K)
(6.1--7.1)$\times$10$^{-20}$~g~cm$^{-3}$. Then, with the DCF method, the magnetic field strength is estimated to be 34--76~$\mu$G. 
If we adopt the formulation in \cite{2021Skalidis}, which considers
%provides better estimates than the DCF method because it incorporates 
compressible and magnetized turbulence: %Unlike the previous method, which provides a more accurate estimate of the magnetic field strength in a compressible and magnetized turbulence environment,
\begin{equation}
|\boldsymbol{B}_\mathrm{pos}| = \sqrt[]{2\pi\rho_\mathrm{gas}}\frac{\sigma_{\mathrm{NT}}}{\sqrt[]{\sigma_\theta}},
\end{equation}
the magnetic field strength in the dense core is estimated to be 23--37~$\mu$G. %\cite{2003Vallee} estimated the magnetic field strength of $\sim$150 $\mu$G over the central $\sim$40$\arcsec$ region, which is one order of magnitude larger than our estimation. 

%These magnetic field ranges include the angular dispersion and the mean volume density. 

The dimensionless mass-to-flux ratio ($\mu_\mathrm{pos}$) of the dense core on the plane of sky is derived as:
\begin{equation}
\mu_\mathrm{pos} = 2\pi\ \sqrt[]{G}\frac{\Sigma}{B_\mathrm{pos}},
\end{equation}
where $G$ is the gravitational constant and $\Sigma$ is the mean surface density \citep{1978Nakano}. $\Sigma$ is calculated from the total enclosed mass, e.g., the core and central protostellar masses, divided by the core area within a box of 60$\arcsec$ from the protostar. The mass-to-flux ratio in the dense core is estimated to be 1.5--3.8 with the DCF method and 3.1-5.7 with the method in \cite{2021Skalidis}, suggesting that the dense core of IRAS~16544 is supercritical. 
%The transcritical mass-to-flux ratio suggest that the magnetic field is comparable to gravitational collapse.
%Note that, the uncertainties in the magnetic field strength and mass-to-flux ratio were computed with the propagation of errors of the magnetic field angular dispersion and density.

\subsection{Infalling streamers observed with ALMA} \label{subsec:streamers by ALMA}
\subsubsection{Spatial and velocity structures of streamers} \label{subsubsec:identified_streamers}
\begin{figure*}[t]
\centering
\includegraphics[width=180mm, angle=0]{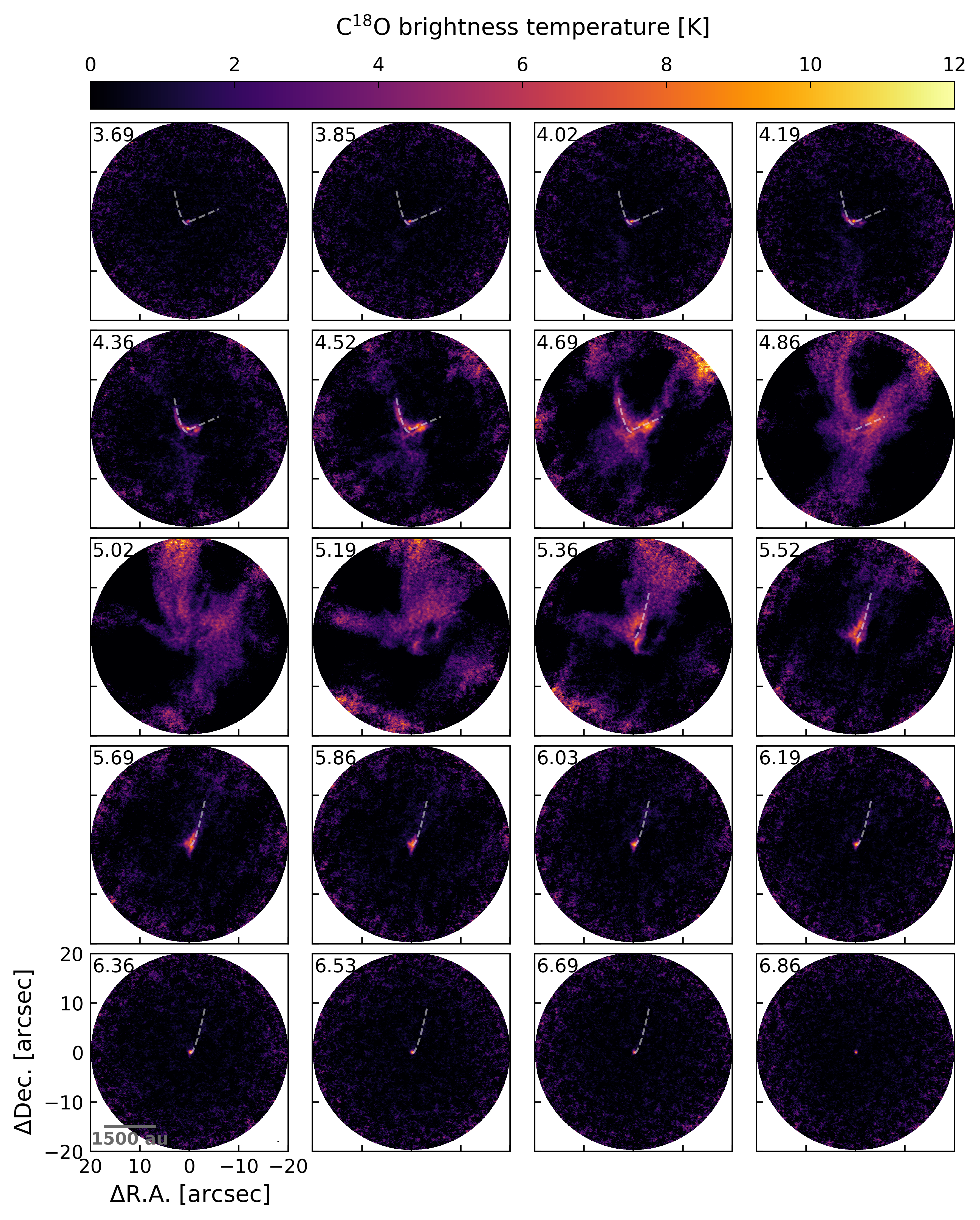}\\
\caption{Velocity channel maps of the C$^{18}$O (2--1) emission observed with ALMA. The maps are generated by combining the 12-m and 7-m array data. The LSR velocity of each channel in km s$^{-1}$ is indicated in the upper-left corner of each panel. White dashed lines indicate the model trajectories of the NE, NW, and NNW streamers, respectively. An ellipse at the bottom right corner in the lower-left panel denotes the synthesized beam size and the rms noise level is 0.52 K.
%The systemic velocity is 5.0 km s$^{-1}$.
}
\label{fig_channelmap}
\end{figure*}
Figure~\ref{fig_channelmap} shows the C$^{18}$O velocity channel maps generated from the combined ALMA-only data. The peak brightness temperature within the radius of 15$\arcsec$ is 12~K. Thus, the C$^{18}$O emission is most likely optically thin. %The dusty disk is elongated along the northeast to the southwest, and 
The C$^{18}$O emission associated with the disk is detected at $V_\mathrm{LSR}$ $<$ 4.02~km~s$^{-1}$ and $V_\mathrm{LSR}$ $>$ 6.36~km~s$^{-1}$, as described in detail in \cite{2023Kido}. Here, we focus on the extended structures at the lower velocities.

%Four elongated components with lengths of greater than 1500 au are detected in total: one in the northeastern direction at 4.19--5.02 km s$^{-1}$, another in the northwestern direction at 4.36--5.02 km s$^{-1}$, one in the eastern direction at 5.02--5.86 km s$^{-1}$, and one more in the northeastern direction at 5.36--6.19 km s$^{-1}$. 
Several elongated structures with lengths of 2$\arcsec$ to 20$\arcsec$ are detected around the protostar in the velocity channel maps. The three most prominent ones with peak intensities of 11--12~K are observed in the northeast at $V_\mathrm{LSR}$ $=$ 4.19--5.02~km~s$^{-1}$, in the northwest at $V_\mathrm{LSR}$ $=$ 4.36--5.02~km~s$^{-1}$, and in the north-northwest at $V_\mathrm{LSR}$ $=$ 5.19--6.03~km~s$^{-1}$, (see Figure~\ref{fig_channelmap}).
%Similar structures can be seen in the higher angular resolution data ($\theta$ $\sim$ 0$\farcs$24) \citep{2023Kido}, although they are slightly stronger emission by --$\%$ and more extended to the north. 
%The detected velocity regimes are also same as that of \cite{2023Kido}, NE streamer and the NW blueshifted component have the blueshifted velocity range of 4.19-4.86 km s$^{-1}$ and the NNW redshifted component has the redshifted velocity regime of 5.19-5.86 km s$^{-1}$, respectively. 
These elongated structures are unlikely to be associated with outflows because their velocity structures are different from those of typical outflows \citep[e.g.,][]{2007Arce, 2024Lopez-Vazquez}. The emission at the higher velocities is detected in the vicinity of the protostar, while that at the lower velocities is farther away from the protostar.
Furthermore, the blueshifted outflow is observed on the southeastern side (P.A. $\sim$ 135$\degr$), while the redshifted outflow is on the northwestern side (P.A. $\sim$ $-$17$\degr$). The two blueshifted elongated structures are in the north, opposite to the direction of the velocity gradient of the blueshifted outflow \citep{2023Kido}. 
%In addition, the $^{12}$CO emission, primarily tracing the outflow, is detected at redshifted velocities of $V_\mathrm{LSR}$ $=$ 5--20~km ~s$^{-1}$, much higher than the velocities of the redshifted elongated structure. 
%\textbf{Nevertheless, the contamination from the outflow emission in the redshifted elongated structure cannot be ruled out, and this does not affect our conclusion in the present paper, as discussed below.}
On the other hand, both the redshifted elongated structure and the redshifted outflow are located to the northwest, and there is a slight overlap of the velocities between the redshifted outflow ($V_\mathrm{LSR}$ $=$ 5--20~km~s$^{-1}$) and the redshifted elongated structure. %the velocity range of the redshifted component is already mentioned in line 346 
Thus, in contrast to the two blueshifted elongated structures, it is possible that the redshifted elongated structure traces parts of the molecular outflow, however, this does not affect our conclusion in the present paper, as discussed below.
On the contrary, as we show below, the detailed spatial and velocity structures suggest that the nature of the redshifted elongated structure is the same as those of the two blueshifted elongated structures.
%Therefore, these elongated structures are unlikely associated with outflow in IRAS~16544.

Figure~\ref{fig_almamom0}(a) and (b) show the moment 0 maps of the C$^{18}$O emission at the blue- and redshifted velocities, respectively. In the moment 0 maps, the three most prominent elongated structures are clearly seen in the northeast (NE), northwest (NW), and north-northwest (NNW) regions. The blueshifted NE elongated structure has been suggested to be an infalling streamer by \cite{2023Kido} (hereafter NE streamer). %, and other two components are also similar to NE streamer.
Below we examine the spatial and velocity structures of the other two elongated structures in comparison with those of the NE streamer.

%The spatial and velocity structures of the elongated components are analyzed based on the infall-rotating envelope model in \cite{1976Ulrich} and \cite{1981Cassen}, which is often called the UCM model and adopted to describe the kinematics of infalling material \citep[e.g.,][]{2014Yen, 2020Pineda, 2023Flores, 2023Aso, 2024Lin, 2022Thieme}. This model assumes that a particle falls toward a point mass with a conserved angular momentum and zero initial energy. 
Blue and red lines in Figure~\ref{fig_almamom0} represent the free-fall trajectories coincident with the elongated structures. These trajectories are derived with the infall-rotating envelope model in \cite{1976Ulrich} and \cite{1981Cassen}, which is often called the UCM model and adopted to describe the kinematics of infalling material \citep[e.g.,][]{2014Yen, 2020Pineda, 2023Flores, 2023Aso, 2024Lin, 2022Thieme}. This model assumes that a particle falls toward a point mass with a conserved angular momentum and zero initial energy. 
%that an axisymmetric envelope falls to the centrifugal radius ($r_\mathrm{d}$) under the gravitational potential of the central star, while conserving angular momentum. The spatial and velocity components in the polar, azimuthal, and radius directions are described in \cite{1976Ulrich} and \cite{1981Cassen}, 
The model has 6 free parameters: stellar mass $M_*$, centrifugal radius $r_\mathrm{d}$, the initial polar angle $\theta_0$ and azimuthal angle $\phi_0$ of the particle, and two parameters $i_\mathrm{s}$ and $\theta_\mathrm{s}$ for the direction of the rotational axis. 
The spatial and velocity structures of the NE streamer have been successfully reproduced with this model in \cite{2023Kido} obtaining the values of $M_*$ $=$ 0.14~$M_\odot$, $r_\mathrm{d}$ $=$ 100~au, $\theta_0$ $=$ 90$\degr$, $\phi_0$ $=$ 64$\degr$, \textit{i}$_\mathrm{s}$ $=$ 73$\degr$, and $\theta_\mathrm{s}$ $=$ 60$\degr$. 

To compare the morphologies of the NW and NNW elongated structures with the model infalling trajectories, we assume that their rotational axes are aligned with the disk and that their centrifugal radii are the same as the gas disk radius. Thus, we fix \textit{i}$_\mathrm{s}$ $=$ 107$\degr$, $\theta_\mathrm{s}$ $=$ $-$45$\degr$, $r_\mathrm{d}$ $=$ 55~au, and $M_*$ $=$ 0.14~$M_\odot$, as measured by \cite{2023Kido}.
%There remains only three free parameters, initial position polar and azimuthul angle and the radius at the initial position for particle. Then, the observations can be replicated for the NW component in Figure \ref{fig_almamom0}(a) with $\theta_0$ $=$ 20$\degr$ and $\phi_0$ $=$ 170$\degr$, and for the NNW component in Figure \ref{fig_almamom0}(b) with $\theta_0$ $=$ 40$\degr$, and $\phi_0$ $=$ 50$\degr$. 
There remains only two free parameters: $\theta_0$ and $\phi_0$. We visually inspect and compare the observed structures with the model trajectories of all combinations of the parameters, and find that the model trajectory with $\theta_0$ of 20$\degr$ and $\phi_0$ of $-$10$\degr$ can match the NW elongated structure and that with $\theta_0$ of 70$\degr$ and $\phi_0$ of $-$120$\degr$ can match the NNW elongated structure, as shown in Figure~\ref{fig_almamom0}(a) and (b). 
We note that there are other parameter combination that can also explain the observed emissions and we discuss the degeneracy of the parameters in Section~\ref{subsubsec:dis_Degeneracy}.

%Next, we compare the velocity gradients of each UCM model with observations. 
Figure~\ref{fig_streamerpv} presents the Position--Velocity (PV) diagrams of the elongated structures cut along the model infalling trajectories. %The observations and UCM models are shown as color scale and white curves. As seen in the PV diagrams,
The model velocity structures (white lines) follow the observed emission (color scale), with the emission at higher velocities becoming more dominant closer to the protostar.
These velocity structures are consistent with the UCM model. 
Note that the bright components seen at the highest redshifted velocity in the PV diagrams for the NE and NW structures and at the highest blueshifted velocity in that for the NNW structure are smoothly connected to the Keplerian components associated with the disk.
%Although there are offsets between models and observations, they are mostly consistent. 
%%In addition to the spatial structures, the velocity structures of the NW and NNW elongated structures also match the expectation from the free-fall material, as shown in the PV diagram. 
%can be explained with the streamer model, and 
%Thus, the NW and NNW elongated structures are likely accretion streamers, similar to the NE elongated structure.
 
\subsubsection{Mass infalling rates of streamers} \label{subsubsec:streamer properties}
\begin{figure*}[t]
    \centering
    \includegraphics[width=180mm]{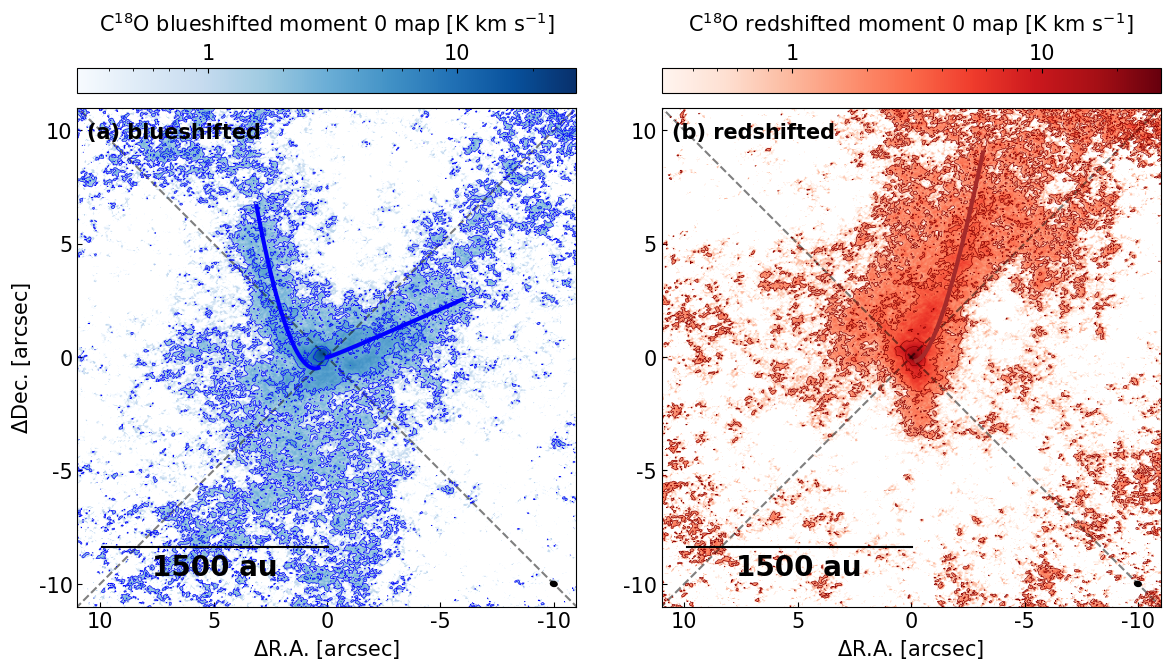}
    \caption{ALMA 12-m and 7-m array combined C$^{18}$O ($J$ $=$ 2--1) blue- (a) and red-shifted (b) moment 0 maps. The integrated velocity ranges are $V_\mathrm{LSR}$ $=$ 1.2--4.9~km~s$^{-1}$ and $V_\mathrm{LSR}$ $=$ 5.2--8.9~km~s$^{-1}$, respectively. The dashed lines from north-east to south-west and from north-west to south-east indicate the disk major and minor axes, respectively. 
    The solid curves denote the infalling trajectories derived from the UCM model that can match the observations.
    Contours are drawn at 3, 7, 15, and 25$\sigma$ levels, where 1$\sigma$ $=$ 0.38~K~km s$^{-1}$.
    Black filled ellipses at the bottom-right corner show the synthesized beams of the C$^{18}$O data.
    }
    \label{fig_almamom0}
\end{figure*}
\begin{figure*}[t]
    \centering
    \includegraphics[width=180mm]{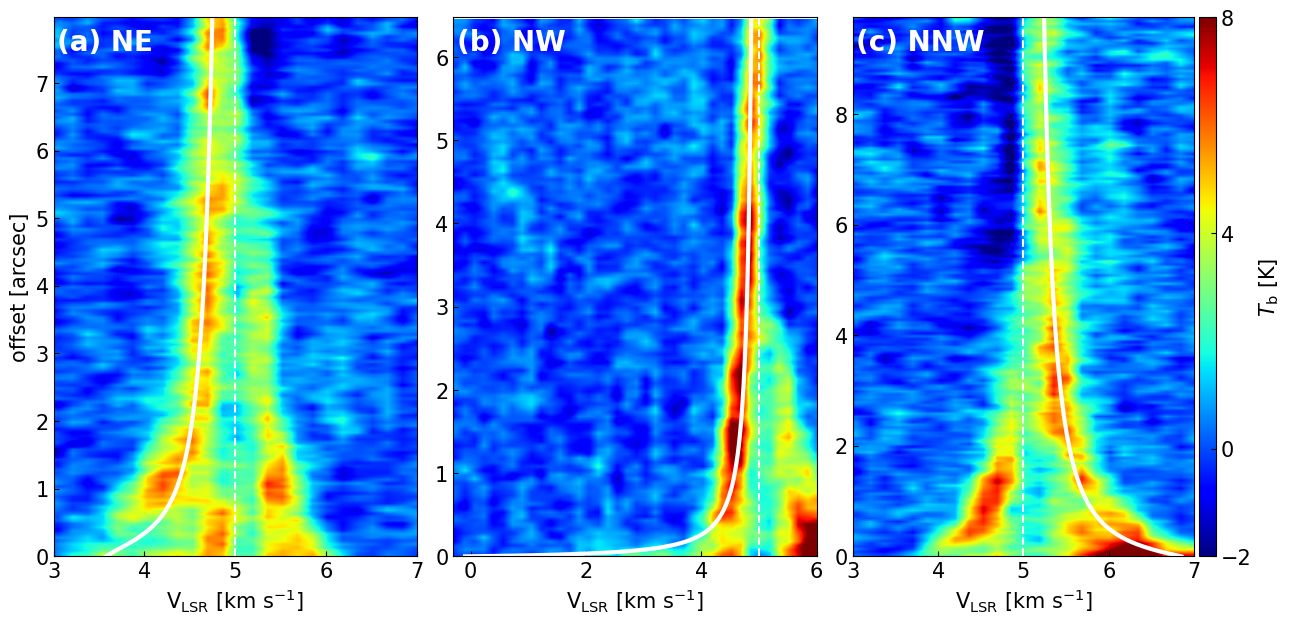}
    \caption{PV diagrams of the C$^{18}$O emission along each model trajectory shown in Figure~\ref{fig_almamom0}. White curves show the velocity structures of the UCM models. The vertical dashed lines indicate the systemic velocity of 5.0~km~s$^{-1}$.}
    \label{fig_streamerpv}
\end{figure*}
%Figure \ref{fig_almamom0} shows the blue- and red-shifted moment 0 maps of ALMA combine  data in C$^{18}$O ($J$ = 2--1) emission. The line morphology is consistent with that observed in ALMA LP eDisk \citep{2023Kido}, while for the total-flux, combined data is - times larger than the eDisk-only. We can find a compact, strong emission associated with the protostellar disk, along with a diffuse component extending to the northern and southern region, same as \cite{2023Kido}.
%To derive the physical properties of streamers, such as mass and length, it is necessary to extract them using some kind of algorithm. Following \cite{2020Hsieh} and \cite{2022Thieme}, we attempted to pick up theese streamers in position-position-velocity space using Python package \textit{astrodendro} \citep{2008Rosolowsky}. However, due to the low S/N ratio in these regions, the streamers were recognized as multiple compact components. To prevent this, we adjusted the `npix' parameter, which is the minimum number of pixels of the masked region. Despite this adjustment, the mask did not extend to the edges of streamers and were truncated partway. Additionally, the two blueshifted streamers, which have the same velocity range, were identified as a single component by dendrogram analysis, making it impossible to distinguish between them. Hence, in this paper, we stopped to pick out the streamers using algorithm. 
%In this section, we estimate the length, width, column density, mass, free-fall time, and infalling rate of streamers. 
The comparison with the UCM model suggests that the three elongated structures (NE, NW, and NNW) likely trace infalling streamers.
We set the 3$\sigma$ emissions in the moment 0 maps (Figure~\ref{fig_almamom0}) as the edge of the streamers.
%%%To estimate the mass and mass infall rates of streamers, we identify their starting points and widths.
%The edge the streamers with above 3$\sigma$ emissions in the moment 0 maps are used as starting points, and the widths are determined based on the model trajectories to cover the 3$\sigma$ emissions in moment 0 maps, so there is no issue even if we cannot identify with algorithm.
%%%The edge the streamers with above 3$\sigma$ emissions in the moment 0 maps are used as starting points.
%%%Assuming accretion from the edge of the FoV, we compared the PV diagrams with models and found that the emission overlapping with the orbits can be explained by the infalling streamers. 
%%%However, since the emission is below 3$\sigma$ in the moment 0 map, we adopt the location with emission above 3$\sigma$ to define the starting points of streamers. 
%%%
%We estimated the edge of the 3$\sigma$ emission as lengths of streamers based on Fig.~~\ref{fig_almamom0}, and then derived its velocity ranges. They are 3.6--4.7 km s$^{-1}$, $-$0.1--4.9 km s$^{-1}$, and 5.2--6.8 km s$^{-1}$ for 
The NE, NW, and NNW streamers are detected at velocities of $V_\mathrm{LSR}$ $=$ 3.6--4.7~km~s$^{-1}$, $-$0.1--4.9~km~s$^{-1}$, and 5.2--6.8~km~s$^{-1}$, respectively, as shown in their PV diagrams. For each streamer, we generate a moment 0 map integrating over its velocity range (see first column of Figure \ref{fig_different_model}). 
We estimate its width covering the emission above the 3$\sigma$ level in its moment 0 map. 
The widths of the three streamers are estimated to be 450 to 900~au,
%3$\arcsec$ to 6$\arcsec$, 
and the lengths are estimate to be $\sim$1300~au on the plane of the sky. 
%%%We derived the velocity regimes of the model trajectories and they are 3.6--4.7 km s$^{-1}$, $-$0.1--4.9 km s$^{-1}$, and 5.2--6.8 km s$^{-1}$ for NE, NW, and NNW streamers.
%We derived the velocity regimes of streamers in Sec.~\ref{subsubsec:channel map of streamers}, and can create moment 0 maps for each streamer within these velocity ranges. However, since models have degeneracy and the their velocity ranges differ between the models, we used the moment 0 maps, integrating channels of $\pm$5.0 km s$^{-1}$ from the systemic velocity. overestimate???
%then identified the widths of streamers. They are determined based on the model trajectories to cover the 3$\sigma$ emissions in moment 0 maps, so there is no issue even if we cannot identify with algorithm.
Then we calculate the integrated C$^{18}$O flux of the streamer in its moment 0 map and estimated its mass with the same method described in Section~\ref{subsubsec:dense core mass}.
%%%Using all pixels within the regions defined above, we calculated the mass of the streamers following the same method described in Section~\ref{subsubsec:dense core mass}.
%We then use all pixels within these defined regions following the same method described in Section \ref{subsubsec:dense core mass}. 
The masses of the NE, NW, and NNW streamers are estimated to be 1.3$^{+1.4}_{-0.2}$$\times$10$^{-3}$, 
2.2$^{+2.1}_{-0.4}$$\times$10$^{-3}$, 
and 4.0$^{+4.0}_{-0.6}$$\times$10$^{-3}$~$M_{\odot}$, respectively. 
The uncertainties originate from the possible ranges of the C$^{18}$O abundance of (1--2)$\times$10$^{-7}$ \citep{2022GOng} and the excitation temperature of 10--30~K.
%We note that the intensity distributions of the NE and NW streamers partially overlap on the plane of the sky. However, the mass in the overlapping region is small and is only 3.4$\times$10$^{-5}$ $M_\odot$. Thus, it does not affect our results and discussions.
%there are overlapped regions to calculate the mass. Their corresponding mass of \textcolor{red}{-- $M_\odot$} is very small, so we do not account for the effect of that.

The lengths of the NE, NW, and NNW streamers in three-dimensional space are estimated to be $\sim$4790~au, $\sim$2830~au, and $\sim$6130~au, respectively, based on the infalling trajectories from the UCM model.
%%The 3D distances corresponding to the starting points defined on the moment 0 maps can be determined by comparison with the UCM model, which allows us to calculate these values.
%The radii of NE, NW, NNW streamers are calculated to be $\sim$4790 au, $\sim$2830 au, and $\sim$5790 au, respectively.
%Assuming free-fall collapse driven by the central star,
Their free-fall timescales are calculated to be 1.0$\times$10$^5$, 4.5$\times$10$^4$, and 1.4$\times$10$^5$~yr from their lengths divided by their free-fall speeds at their lengths ($t_\mathrm{ff}$ $=$ $\sqrt{{R^3} / {2GM_*}}$), and the mass infalling rates are estimated to be  1.3$^{+1.4}_{-0.2}$$\times$10$^{-8}$, 
4.8$^{+4.7}_{-0.7}$$\times$10$^{-8}$, and 2.8$^{+2.7}_{-0.4}$$\times$10$^{-8}$~$M_{\odot}$~yr$^{-1}$ using their estimated masses and infalling timescales for the NE, NW, and NNW streamers, respectively.
%%%Similarly, the specific angular momentum can be derived from the comparison since both centrifugal radius and the starting points are known.

%For the widths, we defined them to cover the 3$\sigma$ emission along the obtained model trajectories, resulting value of 3$\arcsec$, ?$\arcsec$, and ?$\arcsec$.
%Although this estimation method may include the Keplerian rotation component in the gas disk, leading to an overestimate, it is sufficient for the discussion in this paper (see Section \ref{subsec:dis_mass_accretion} for details).
%These lengths are similar with the lengths in moment 0 maps, which means all streamers extend along the plane of sky. 
%2.1, 5.3, and 5.6 x 10$^{-8}$ M$_{\odot}$ yr$^{-1}$
%This is one order of magnitude smaller than a typical mass-infall rate of Class 0/I stages and other streamers.
%The magnitude of the specific angular momentum are $\sim$111, $\sim$28, and $\sim$53 au km s$^{-2}$, respectively. IRAS 16544 has a Keplerian disk of 55 au associated with the protostar of 0.14 $M_\odot$, and it is estimated to be $\sim$83 au km s$^{-1}$. These results indicate that NW and NNW streamers are accreting onto the inner part of the Keplerian disk, while NE streamer is accreting onto the outer part.

%\subsubsection{PV diagrams along the major and minor axes}
\subsection{Infalling protostellar envelope observed with ALMA and JCMT}\label{subsec:model}

\begin{figure*}[t]
\centering
\includegraphics[width=180mm, angle=0]{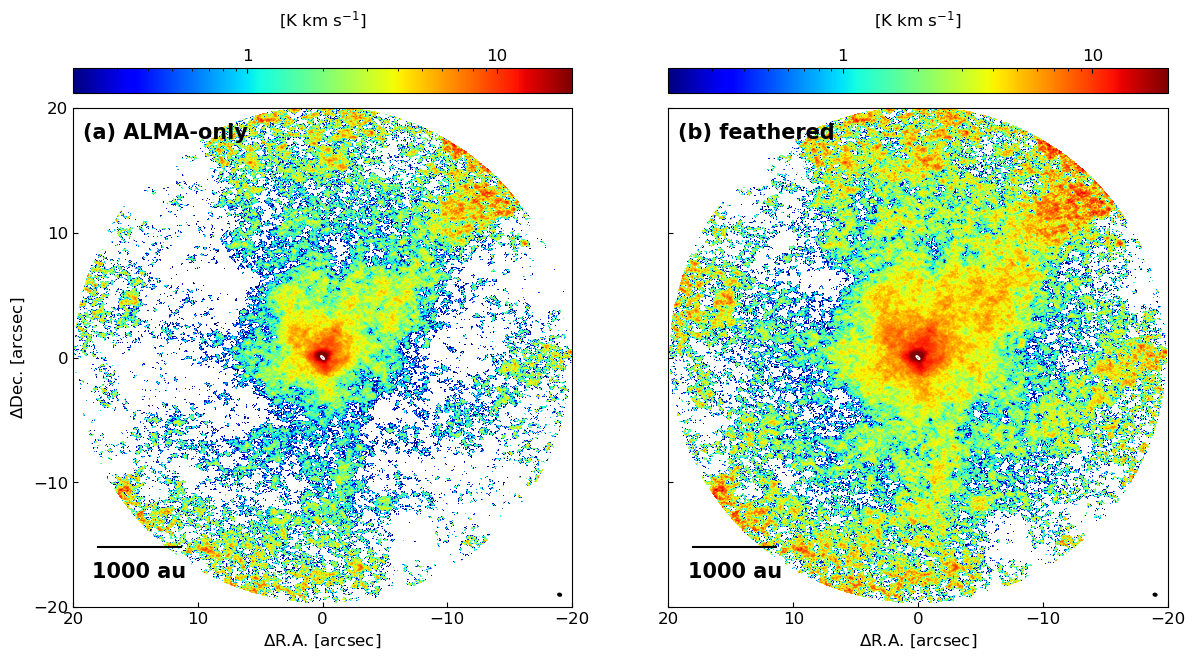}\\
\caption{Comparison of the C$^{18}$O moment 0 maps from the ALMA-only (a) and feathered data (b), integrated over the velocity range of $V_\mathrm{LSR}$ $=$ 1.2--8.9~km~s$^{-1}$. %1.18-8.86
%\textbf{Dashed circles denote the circle with a radius of 10$\arcsec$ from the protostar.}
White contours in both panels show the 10$\sigma$ level of the 1.3-mm continuum emission observed with eDisk \citep{2023Kido}.
Filled black ellipses at the bottom right corners denote the synthesized beams.}
\label{fig_feather_mom0}
\end{figure*}
\begin{comment}
\begin{figure*}[t]
\centering
\includegraphics[width=180mm, angle=0]{./figures/test31_major_pv.png}\\
\includegraphics[width=180mm, angle=0]{./figures/test31_minor_pv.png}\\
%\includegraphics[width=180mm, angle=0]{./test15_major_pv.png}\\
\caption{Left and right columns present the PV diagrams of the C$^{18}$O (2--1) emission from our feather data and kinematical models, respectively. The upper and lower panels are the PV diagrams of the C$^{18}$O (2--1) emission along the disk major axis (P.A. $=$ 45$\degr$) and minor axis (P.A. $=$ 135$\degr$). The blue, red, and green dots denote the positions and velocities where the emission may be from the NE, NW, and NNW streamers, respectively.}
\label{fig_majorpv}
\end{figure*}
\end{comment}

\begin{figure*}[t]
\centering
\includegraphics[width=180mm, angle=0]{
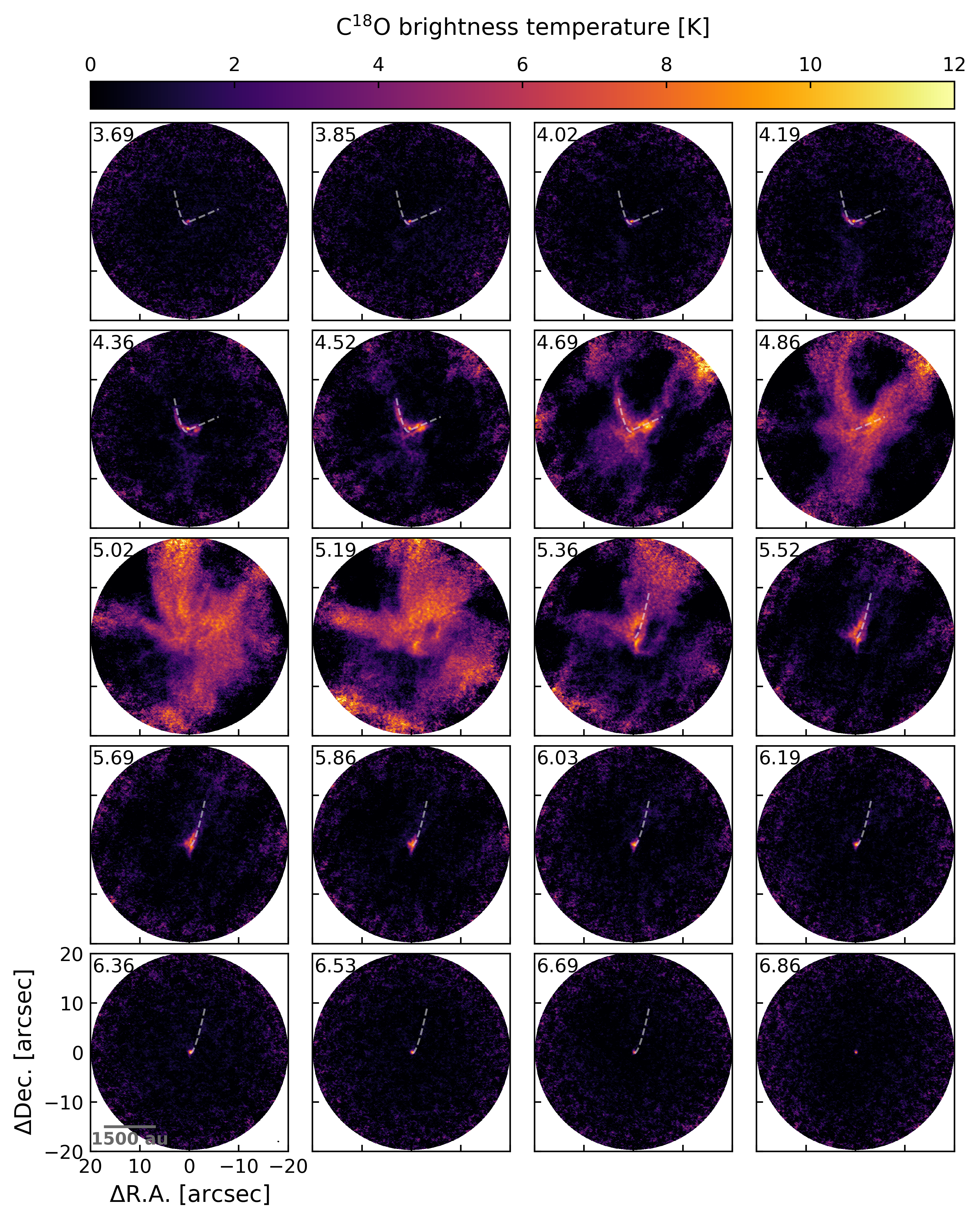
}\\
\caption{Same as Figure~\ref{fig_channelmap}, but for the feathered image.}
\label{fig_channelmap_feather}
\end{figure*}

\begin{figure*}[t]
\centering
\includegraphics[width=180mm, angle=0]{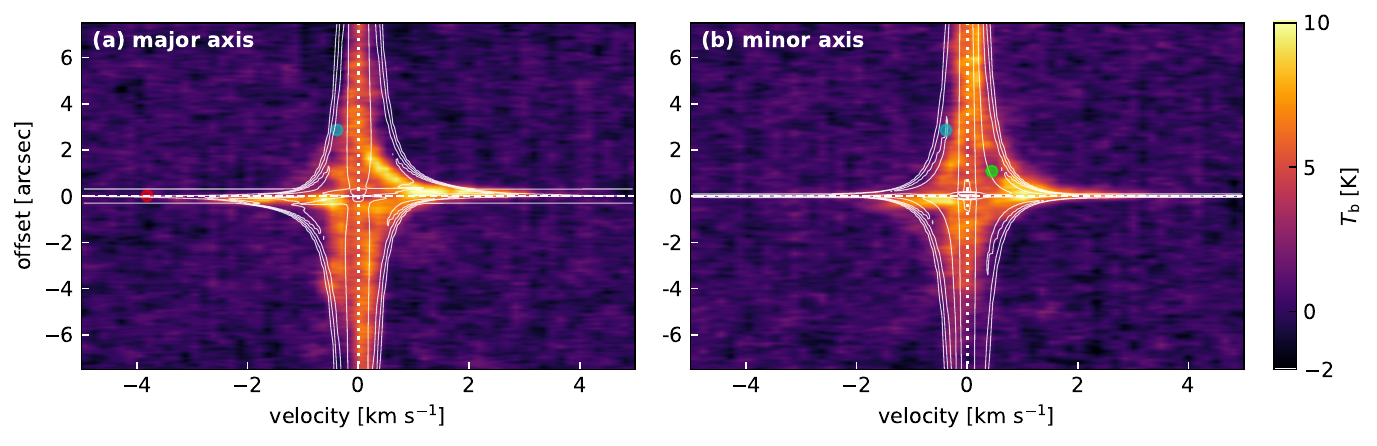}\\
\caption{C$^{18}$O (2--1) feathered (color) and model (white contours) PV diagrams cut along (a) the major (P.A. $=$ 45$\degr$) and (b) minor (P.A. $=$ 135$\degr$) axes of the central disk. (a): The upper side corresponds to the northeastern side along the disk major axis, while the lower side corresponds to the southwestern side. (b): The upper side represents the southeastern side along the disk minor axis, while the lower side represents the northwestern side.
The contour levels are 3, 8, 15, 20, and 25$\sigma$, where 1$\sigma$ is the same value as the observation.  
Cyan, red, and green dots denote the offsets and velocities where the infalling trajectories of the NE, NW and NNW streamers, respectively and PV cuts intersect. 
%The emission at these positions could originate from the NE, NW, and NNW streamers, respectively.
}
\label{fig_majorpv}
\end{figure*}

Figure~\ref{fig_feather_mom0} compares the C$^{18}$O moment 0 maps generated with the ALMA-only (a) and feathered data (b). In addition to the emission extended to the northern side seen in the ALMA-only map, additional extended emission with a roundish distribution around the protostar is enhanced in the feathered map. 
In the velocity channel maps of the feathered data (Figure~\ref{fig_channelmap_feather}), the emission near the systemic velocity is recovered, and prominent extended emission is more visible on the western and southern sides at velocities from 4.86 to 5.36~km~s$^{-1}$, distinct from the narrow elongated structures of the streamers. 
While these extended components tend to have lower brightness temperature ($<$ 10~K) than that of the streamers (11--12~K), their area is significantly larger than that of of the streamers. As seen in the velocity channels at 5.02 and 5.19 km s$^{-1}$, the extended emission is distributed approximately across the entire field of view.
%While the peak brightness temperatures of these extended components within a radius of 10$\arcsec$ ($<$ 10~K) are lower than those of the streamers (11--12~K), the area of the extended components dominates that of the streamers.
%The mean H$_2$ column densities of the streamers and the extended components are calculated to be - cm$^{-2}$ and - cm$^{-2}$, respectively. 
%These extended components have brightness temperatures ranging from 3 to 8 K, with the peak brightness temperature only 50\% lower than those of the streamers of 11--12 K. The area of these extended components dominates that of the three streamers. 
This suggests that the streamers comprise only a portion of the entire protostellar envelope.

The PV diagrams cut along the disk major and minor axes of the C$^{18}$O emission from the feathered image cube are presented in Figure~\ref{fig_majorpv}.
The PV diagrams show a diamond-shaped feature suggestive of a rotating and infalling envelope \citep[e.g.,][]{1997Ohashi, 2014Ohashi, 2014Sakaib, 2016Oya, 2018Takakuwa, 2022Okoda}. 

We construct kinematical models of an infalling and rotating protostellar envelope with an embedded Keplarian disk to compare with the observed PV diagrams. 
Following \cite{2024Takakuwa}, we adopt the description of an infalling and rotating envelope from the UCM model and included an axisymmetric flared disk at the center in our kinematical model. The density profile of the model envelope is scaled to match the core mass and size of IRAS~16544 observed with JCMT. The mass, radius, and orientation of the model disk are adopted to be the same as those in IRAS~16544 observed with ALMA \citep{2023Kido}. For simplicity, the model assumes an isothermal temperature of 15 K as the C$^{18}$O emission in IRAS~16544 is likely optically thin. 
%Our goal is to capture the isotoropic infalling envelope, and incorporating these effects is beyond of the scope of our study.
%We first modeled Keplerian disk, and then protostellar envelope (further details about the model are described in their paper). 
All the model parameters are listed in Table\ref{table:model parameters}. 
%Parameters obtained from previous observations, $d$, $M_*$, $r_\mathrm{d}$, $M_\mathrm{d}$, disk P.A. and $i$, are fixed (parameters are listed in Table\ref{table:model parameters}). 
%The vertical scale height $H(r)$ is calculated with the dust temperature $T_0$ at $r$ $=$ $r_0$. In this model, the dust temperature at $r_0$ $=$ 1 au is set to 100 K, based on the dust peak temperature of eDisk data was approximately 100 K. 
%The dust and gas surface density profile $\Sigma(r)$, the gas mass density $\rho$ and velocity profile of the disk is same as those in \cite{2024Takakuwa}. For the protostellar envelope model, we use the size and mass obtained in Section \ref{subsubsec:dense core mass}, and the density profile adopted is the same as that used by \cite{2024Takakuwa}. The shape of the envelope is assumed to be spherical.
%we calculate the mass-infall rate ($\dot{M}$) from JCMT C$^{18}$O moment 0 map within a radius of 2000 au and used a density profile that matches this $\dot{M}$. The other structures, velocity vectors and the degree of flattening of the envelope are same as those in \cite{2024Takakuwa}. 

\begin{table}[t!]
\centering
\caption{Parameters adopted in the protostellar disk+envelope model.}
    \begin{tabular}{lc}
    \hline\hline
    \colhead{Parameter} & \colhead{Value}\\ \hline
    radial range & 1-9000~au\\
    distance &  151~pc\\
    stellar mass & 0.14~$M_\odot$ \\
    %bolometric luminosity & 0.89 L$_\odot$ \\
    disk radius & 55~au \\
    disk mass & 1.0$\times$10$^{-2}$~$M_\odot$ \\
    disk P.A. & 45$\degr$ \\
    disk inclination & 73$\degr$ \\
    envelope mass & 0.52~$M_\odot$\\
    C$^{18}$O abundance & 1.7$\times$10$^{-7}$\\
    \hline
    \end{tabular}
\label{table:model parameters}
\end{table}

We then perform radiative transfer calculations using RADMC-3D \citep{2012Dullemond} and generate a model C$^{18}$O image cube of our kinematical model. 
Our model adopts the dust opacity table in \cite{2003Semenov} and the C$^{18}$O abundance of 1.7$\times$10$^{-7}$, assuming LTE condition.
%which is the smaller dust particles compared to DSHARP opacity is used in our model.  
%The calculation of the radiative transfer of the C$^{18}$O line is conducted under the assumption of LTE and the C$^{18}$O abundance of 1.7 $\times$ 10$^{-7}$. 
The model images are convolved with a Gaussian beam of the same size as that of our feathered map of the C$^{18}$O emission. 
The PV diagrams cut along the disk major and minor axes extracted from the model image cube are overlaid on the observed ones in Figure~\ref{fig_majorpv}. 
The observed PV diagrams indeed show similar velocity structures to those of the isotropic infalling and rotating envelope model. %, although the intensity distributions of the observations and model do not fully match. 
Along the major axis, both diagrams show tilted, diamond-shaped features.
%In addition, strong Keplerian rotating components are visible in the first and third quadrants. 
%Regarding Figure \ref{fig_minorpv}, the model shows a typical infalling motion with an asymmetric diamond-shaped PV diagram, reproducing the observed structure.
Along the minor axis, infalling motion is expected to induce a clear velocity gradient, where the northern part is more redshifted with fainter blueshifted emission and the southern part is more blueshifted with fainter redshifted emission, as seen in the model PV diagram. 
This feature is also seen in the observed PV diagram. 
%This velocity feature is observed in the PV diagram of the C$^{18}$O emission in IRAS~16544.

We note that the PV cuts and the trajectories of the streamers intersect. 
The positions and velocities of these intersections are labeled with the colored dots in the PV diagrams. 
There is only a minor contribution from the streamers in the PV diagrams. 
%While the accretion streamers described above intersect at the spatial and velocity points as shown in the colored dots, the rotating and infalling envelope shown in the P-V diagrams is distinct from those accretion streamers.
%Although these PV cuts intersect with the infalling streamers, the contribution from the streamers in the PV diagrams is negligible. The dots in the PV diagrams indicate the positions and velocities where the emission may be from the streamers. 
The majority of the emission in the PV diagrams is from the extended envelope and shows the feature of infalling motion. 
%while the components that match the other parts of the model are considered to be part of the infall envelope.
The good agreement between the observed and the model velocity structures in the PV diagrams suggests that the extended envelope observed in the feathered map is also infalling. Meanwhile, in addition to the detection of streamers, the observed intensity distribution cannot be fully explained by our axisymmetric envelope model, so that the density distribution of the protostellar envelope around IRAS~16544 is probably not axisymmetric. 

\subsection{Influence of the Missing Flux}
\label{subsubsec:missing flux}
\begin{figure}[t]
    \centering
    \includegraphics[width=\columnwidth]{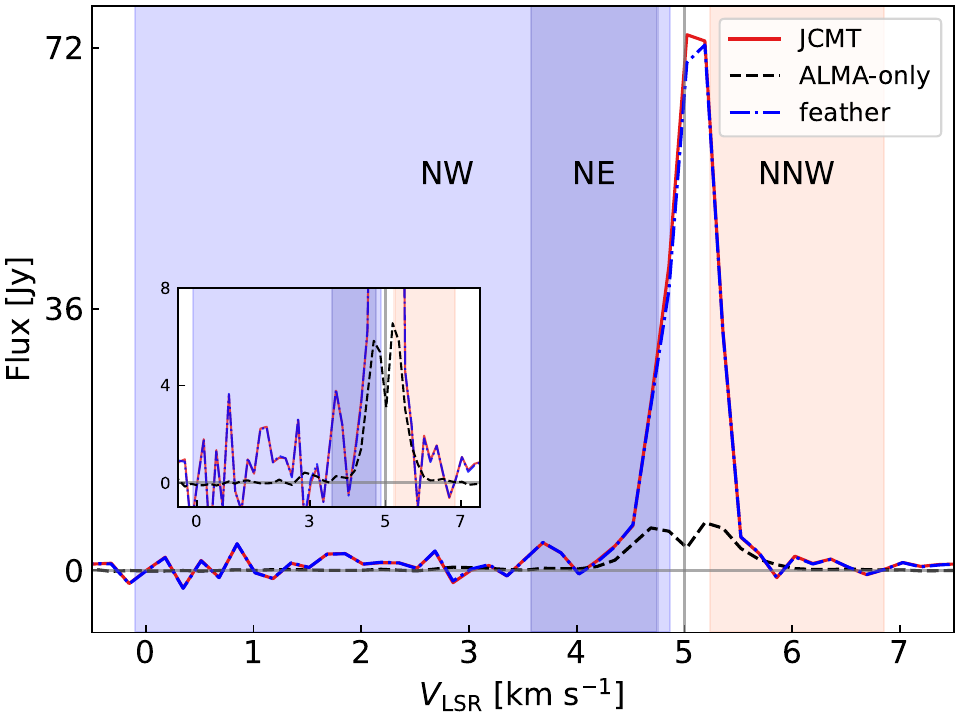}
    \caption{C$^{18}$O ($J$ $=$ 2--1) spectra integrated over the entire ALMA FOV extracted from the JCMT, ALMA-only, and feathered image cubes without primary beam correction. The vertical gray solid line corresponds to the systemic velocity. Shaded regions show the velocity ranges, where the NE, NW, and NNW streamers are observed in the ALMA-only images.
    The insert to the left shows the zoom-in view of the spectra along the vertical axis.
    %The bottom right panel in this figure shows the flux zoom-in view of the spectra.}
    }
    \label{fig_missingflux}
\end{figure}
%These maps are combined five 12m observations, and hence it should severely suffer from the effect of the missing flux.
%Usually, the extended emission is filtered out, so called missing flux, in interferometer like ALMA due to the lack  of short uv baselines.
%When estimating the mass with 12-m array observations, we should take into account for the effect of the missing flux due to the lack  of short uv baselines. 
%We have characterized the streamers with the ALMA-only data as shown in Section \ref{subsec:streamers by ALMA}. Since the ALMA-only data must have missing flux, it would be important for us to discuss whether the missing flux has any impact on the results based on the ALMA-only data. To assess the possible bias due to the missing flux, we compare the fluxes measured in the JCMT, ALMA-only, and feathered data.  
Although all three streamers have widths smaller than the MRS of 6$\farcs$2 as shown in the first column of Figure~\ref{fig_different_model}, we assess the possible bias due to the missing flux in the ALMA-only data by comparing the fluxes measured in the JCMT, ALMA-only, and feathered maps. 
%It would be important for us to discuss whether the missing flux has any impact on the results based on the ALMA-only data.
For this purpose, we regridded the velocity axis of the JCMT data to match that of the ALMA data using \texttt{imregrid} and then measured the flux within the ALMA field of view (FOV). Figure~\ref{fig_missingflux} shows the comparison of the C$^{18}$O spectra integrated over the ALMA FOV from the JCMT, ALMA-only, and feathered images. 
%The data without primary beam correction are used in this figure to avoid negative signals. 
The spectrum of the feathered image coincides with that of the JCMT-only image, showing that the flux is properly recovered with the feathered procedure. 
Significant missing flux of more than 90\% is seen within 0.2~km~s$^{-1}$ from $V_\mathrm{sys}$ in the ALMA-only data, 
and the missing flux reduces to 65\% at the velocities within 1.0~km~s$^{-1}$ from $V_\mathrm{sys}$.
%and the missing fluxes in the velocity ranges of the streamers are estimated to be $\sim$76\%. (or the missing flux within $\pm$1.0 km s$^{-1}$ from V$_\mathrm{sys}$, which all streamers have, is 65\%.)
%NE(3.58-4.75): 80%, NW(-0.1 - 4.87): 220%, NNW(5.24-6.85): 85%
%At velocities of $V_\mathrm{LSR}$ $<$ 4.0 km s$^{-1}$ and 6.0 km s$^{-1}$ $<$ $V_\mathrm{LSR}$, where the compact emission associated with the protostellar disk is seen in the velocity channel maps, the ALMA-only flux is consistent with JCMT data. In contrast, at velocities of 4.0 km s$^{-1}$ $<$ $V_\mathrm{LSR}$ $<$ 6.0 km s$^{-1}$, where extended emission and streamers are seen in the velocity channel maps, the effect of missing flux is significant.
%%%%
%The mean missing flux rates are calculated to be 78$\%$ from 4.0 to 4.9 km s$^{-1}$, 95$\%$ at $V_\mathrm{sys}$, and 81$\%$ from 5.2 to 6.0 km s$^{-1}$, respectively.
%All streamers have the velocity regime of $|$$V_\mathrm{sys}$$-$$V_\mathrm{LSR}$$|$ $<$ 1.0 km s$^{-1}$, which means they might be affected by the missing flux.
%The mean missing flux rates in these streamers for the ALMA data with and without including 7-m array are 
%$\sim$80$\%$, ?$\%$, and ?$\%$, % with 7-m array
%and $\sim$80$\%$, ?$\%$, and ?$\%$, respectively.  % without 7-m array 
To assess the influence of this missing flux on our estimate of the streamer mass using the ALMA-only data, we also calculated the mass from the C$^{18}$O flux in the same area and velocity ranges using the feathered data, and the estimated mass becomes 20\% to 50\% higher than the ALMA-only data. 
%%To quantitative the effect of streamers, with all combined feather image, we estimated the mass and mass infall rate of streamers using same trajectories, widths, and lengths. 
%%%They are estimated to be 1.5$^{+1.5}_{-0.2}$, 3.0$^{+2.9}_{-0.1}$, and 5.4$^{+5.6}_{-0.8}$$\times$10$^{-3}$ $M_{\odot}$, and 1.5$^{+1.5}_{-0.2}$, 6.6$^{+6.4}_{-0.2}$, and 4.1$^{+4.0}_{-0.5}$$\times$10$^{-8}$ $M_{\odot}$ yr$^{-1}$.
%or
%The flux measured in the feather images increased by 15, 36, and 44\% compared to the ALMA-only images?
Thus, the influence of the missing flux on the streamers is limited.
%only moderate
Furthermore, the masses estimated using the ALMA data with and without the 7-m array data are almost identical (see Appendix~\ref{appndix:channelmap}), which suggests that the missing flux in the ALMA-only data can be primarily attributed to filtering out the extended diffuse envelope component. 
%Even after adding the JCMT data, the mass of the streamers remains largely unchanged. This suggests that the C$^{18}$O emission observed by JCMT mainly originates from the diffuse envelope component.
Therefore, in the subsequent discussion, we adopt the streamer mass estimated with the ALMA-only data to avoid the contamination from the extended diffuse envelope component.

%All three streamers extend beyond 10$\arcsec$ as shown in Figure \ref{fig_channelmap}, hence, the calculated mass is underestimated. 
%since the JCMT has a beam size of $\sim$22$\arcsec$, while the resolution of ALMA is 0$\farcs$3.

%Figure \ref{fig_missingflux} shows the comparison of the C$^{18}$O spectra at the stellar position at the angular resolution of 21$\arcsec$ from the ALMA and JCMT data.
%For the velocity range of $|$V$_{sys}$-V$_{LSR}$$|$ $\ge$ 1 km s$^{-1}$ which traces compact emission, missing flux is barely noticeable. Conversely, for the range of $|$V$_{sys}$-V$_{LSR}$$|$ $<$ 1 km s$^{-1}$ which traces extended emission, the effect of missing flux is significantly appearance. We measured the percentage of missing flux for these velocity ranges between two data.
%For the velocity range of $|$V$_{sys}$$-$V$_{LSR}$$|$ $\ge$ 1.0 km s$^{-1}$ which traces compact emission, missing flux is quantitative. Conversely, 
%We calculated recoverable flux rate of ALMA within these velocity regime, so that ALMA only recover $\sim$48\% in the blueshifted emission and $\sim$16\% in the redshifted emission. 
%All streamers have the velocity regime of $|$V$_{sys}$$-$V$_{LSR}$$|$ $<$ 1.0 km s$^{-1}$, which means there is a diffuse extended emission in addition to the streamers.
%the streamer flux would not be completely covered. 
%The average missing flux rate is $\sim$44\% for blueshifted emission and $\sim$79\% for redshifted emission. This indicates that the majority of C$^{18}$O emission comes from the diffuse gas?. 

\section{Discussion}  \label{sec:discussion}
\subsection{Degeneracy of the UCM model}\label{subsubsec:dis_Degeneracy}
In our analysis of the infalling trajectories of the NW and NNW streamers in Section~\ref{subsubsec:identified_streamers}, their centrifugal radii and rotational axes are assumed to be the same as the radius and the rotational axis of the gas disk, and the initial position of the infalling material is estimated to be ($\theta_0$, $\phi_0$) $=$ (20$\degr$, $-$10$\degr$) and (70$\degr$, $-$120$\degr$), respectively.
%(the third and fifth rows in Table~\ref{table:different model}). 
%These values of trajectories are chosen to pass the center of the elongated structures by eye because the widths are determined based on these trajectories. 
We note that these solutions are not unique because of the projection effect.
%In case that we fixed the rotational axis to be the same as that of the disk, 
Changing $\theta_0$ and $\phi_0$ by 10$\degr$ to 40$\degr$ can still result in other possible solutions that have spatial and velocity structures similar to the observations. %of the streamers as those model trajectories described in Sec.~\ref{subsubsec:identified_streamers}. 
%We searched for other possible trajectories that could reproduce the NW and NNW streamers by changing $\theta_0$ and $\phi_0$ by 10 degrees.
For example, the models with $\theta_0$ $=$ 30$\degr$ and $\phi_0$ $=$ 30$\degr$, and $\theta_0$ $=$ 40$\degr$ and $\phi_0$ $=$ $-$140$\degr$ can also reproduce the NW and NNW streamers, respectively. 
%Although the radii and velocity ranges of these possible models are different from those of the original model trajectories, they all have similar mass infall rates with difference within a factor of 4 or an even smaller mass infall rate. 
%Since these two streamers are detected northern side toward the disk, they are basically expected to have $\theta_0$$\leq$90$\degr$. To reproduce the lengths on the plane-of-sky, the larger $\theta_0$ requires a longer radii.
%Because the polar angles of two models are relatively smaller, which means the material accretes from the top?, other possible model trajectories can be reproduced by changing the polar angles from those of the original models within $\phi_0$ $=$ $\pm$20$\degr$, and $\pm$1$\degr$, respectively. Even if we change $\phi_0$, the 3D lengths remain the same as the value in Sec.~\ref{subsubsec:streamer properties} because they do not affect the 3D lengths. The velocity ranges of model streamers are slightly changed, and we expect the mass estimations are also affected. However, the error in streamer mass is within --$\%$.

%When explaining two streamers, NW and NNW, with the UCM model, we assumed that the rotational axis of the envelope ($i_s$ and $\theta_s$) is the same as that of the disk, and that the centrifugal radius ($r_d$) corresponds to the radius of the gas disk. As a result, the only free parameters are the initial position of the particle ($\theta_0$ and $\phi_0$). 
If we allow the rotational axes and centrifugal radii of model streamers to differ from those of the gas disk of IRAS~16544, there are more possible infalling models that can match the observed spatial and velocity structures. 
%For example, a comparison between the UCM model with $i_s$ $=$ 120$\degr$, $\theta_s$ $=$ 150$\degr$, $r_d$ $=$ 100 au, $\theta_0$ $=$ 40$\degr$, and $\phi_0$ $=$ 78$\degr$ and observation is shown in Figure \ref{fig_different_model}(g, h).
For example, with the parameters of $i_s$ $=$ 30$\degr$, $\theta_s$ $=$ $-$30$\degr$, $r_d$ $=$ 100~au, $\theta_0$ $=$ 20$\degr$, and $\phi_0$ $=$ 75$\degr$, the model infalling trajectory  projected on the plane of sky and its corresponding line-of-sight velocity can explain the observation of the NNW streamer (Figure~\ref{fig_different_model}(l) and (k)). 
Figure~\ref{fig_different_model} demonstrates such a degeneracy of the models for each of the three streamers and these model parameters are listed as different model parameters in Table~\ref{table:different model} (Appendix \ref{appndix:Alternative model}).
% of different infalling trajectories that can explain the observations.

%shown as Figure \ref{fig_different_model}(k, l). These models also roughly reproduce the spatial and velocity structures of the observation, indicating that gas infall to the disk. 
Although there is degeneracy in the derived infalling trajectories, the mass infalling rates estimated with these different trajectories only vary within a factor of four (Table~\ref{table:different model}). This is because the mass infalling rate primarily depends on the length of a streamer. 
The conservative upper limit of the mass infalling rate can be given by the shortest length, which is the length projected on the plane of the sky, $\sim$1300~au. 
This approach yields mass infalling rates of 1 to 2$\times$10$^{-7}$~$M_\odot$~yr$^{-1}$, which remains an order of magnitude lower than that of the extended diffuse envelope.
%As discussed in Section~\ref{subsec:dis_mass_accretion}, this leads to the mass infalling rates of 1 to 2$\times$10$^{-7}$ $M_\odot$ yr$^{-1}$, which remains an order of magnitude lower than that of the extended diffuse envelope.
Therefore, these uncertainties in the infalling trajectories do not affect our discussion on the contribution to mass accretion by the streamers in IRAS~16544. 
%If the rotational axis and centrifugal radius are also free parameter, 
%In such cases that rotational axis and centrifugal radius are free parameters, many model solutions with varying radii can be derived. Since the mass infall rate primarily depends on the radii of the streamers, it differs between models. However, as discussed in Sec.~\ref{subsec:dis_mass_accretion}, the upper limits are 1 to 2$\times$10$^{-7}$ $M_\odot$ yr$^{-1}$, these values remain an order of magnitude lower than that of the spherical envelope.
%\textcolor{red}{The estimated mass infalling rate depends on the estimated length of the streamers in 3D space. Although there could be degeneracy in the inferred infalling trajectories (see Sec.~\ref{subsubsec:dis_Degeneracy}), we found that the conservative infall rates are calculated using the plane-of-sky lengths of $\sim$1300 au, resulting in a value of 1 to 2 $\times$ 10$^{-7}$ $M_\odot$ yr$^{-1}$, which are considered upper limits.

%Furthermore, variations in $r_\mathrm{d}$ and $\theta_0$ will change the angular momentum of the streamers. However, 
%Additionally, we note that due to these degeneracies in $r_\mathrm{d}$ and $\theta_0$ in the models, the angular momentum of the streamers may not be robustly constrained.
%it is unclear how much angular momentum the streamers have to accrete to the disk.
In the case of the NE streamer, Figure~\ref{fig_different_model}(c) and (d) show that with a rotational axis and a centrifugal radius aligned with the gas disk and the parameters of $\theta_0$ $=$ 70$\degr$ and $\phi_0$ $=$ 73$\degr$,
%(the second row in Table~\ref{table:different model}), 
the model trajectory can reasonably explain the observed spatial and velocity structures at outer radii larger than 400 au. 
However, this model trajectory is different from that adopted in \cite{2023Kido} and does not pass through the curling SO and SiO emissions observed around the gas disk, which possibly traces the accretion shocks due to the collision between the streamer and disk \citep{2023Kido}. 
%Although the streamers can be explained by the model with the same rotational axis direction as the gas disk, they may still have different dynamical properties from the gas disk. 
Thus, observing the distributions of different molecules, such as shock tracers \citep[e.g.,][]{2023Lee}, may help to disentangle different model trajectories.

\subsection{Mass accretion via streamers} \label{subsec:dis_mass_accretion}
%Although several elongated structures are detected in the channel map, three of which are the most prominent in the moment 0 maps and are identified as streamers. 
Three prominent elongated structures are observed in our ALMA data. Their morphologies and velocity structures are consistent with the trajectories and velocities of infalling material toward IRAS~16544. Therefore, these elongated structures most likely trace infalling streamers toward the protostar. 
The total mass of the three streamers is estimated to be $\sim$7.5$\times$10$^{-3}$~$M_\odot$ and is only $\sim$2$\%$ of the total mass of the dense core with a box of 9000 au even after considering the possible effect of missing fluxes. 
%The mass fraction between streamers with the lengths of \textcolor{red}{$\sim$1300} au on the plane-of-sky and dense core with the box of $\sim$9000 au and found that streamers contain \textcolor{red}{1.5$\%$} of the dense core. 
%The lengths of streamer in 3D space are estimated to be 4000 au.
Even if we compare this value to the dense core mass within a radius of 4000 au, which corresponds to the estimated length of the streamers in the 3D space, the mass of the streamers is still only $\sim$6$\%$ of the total core mass. %5.6%
%In addition, a comparison within the circle with the same length as streamers, \textcolor{red}{1300 au, yields 40$\%$}, suggesting that streamers still have a small fraction of total mass. 
Therefore, these streamers only possess a small fraction of the mass reservoir of IRAS~16544. 
We note that there could be more, unidentified streamers existing in the dense core of IRAS~16544. For example, an elongated structure toward the east is also observed at the velocities of $V_\mathrm{LSR}$ $=$ 5.02--5.36~km~s$^{-1}$ in the velocity channel maps (Figure~\ref{fig_channelmap}). However, this emission is much fainter than the three most prominent streamers, so that its mass contribution is likely even smaller.  

%As discussed in Section \ref{subsubsec:streamer properties}, streamers in IRAS 16544 are feeding the gas at the mass infall rate of 2 to 6 $\times$ 10$^{-8}$ $M_\odot$ yr$^{-1}$, which is two orders of magnitude smaller than that of the envelope. 
The mass infalling rates of these streamers are estimated to be 1 to 5$\times$10$^{-8}$~$M_\odot$~yr$^{-1}$ (Section~\ref{subsubsec:streamer properties}). 
These estimates depend on the estimated length of the streamers in the 3D space. Although there could be degeneracy in the inferred infalling trajectories (see Section~\ref{subsubsec:dis_Degeneracy}), we found that the estimated streamer lengths in all cases are comparable, and thus the derived infalling time and mass infalling rate are similar, within a factor of four.
As discussed in Section~\ref{subsubsec:dis_Degeneracy}, a conservative estimate of mass infalling rates of streamers is calculated to be 1 to 2$\times$10$^{-7}$~$M_\odot$~yr$^{-1}$, and these values can be considered as the upper limits.
%For a conservative estimate, we calculated mass infalling rates using the plane-of-sky lengths of $\sim$1300 au of the streamers, which are the lower limit of their lengths in the 3D space.
%This approach yields mass infalling rates of 1 to 2$\times$10$^{-7}$ $M_\odot$ yr$^{-1}$, and these values can be considered as the upper limits.
%These values remain an order of magnitude lower than that of the spherical envelope.
%Protostars have a substantial amount of gas remaining around the disk. 
These estimated mass infalling rates of the streamers are more than an order of magnitude lower than the theoretically expected mass infalling rate of $\sim$2$\times$10$^{-6}$~$M_\odot$~yr$^{-1}$ in a collapsing isothermal and spherical dense core in the conventional model \citep{1977Shu}.

%In the conventional theoretical model of a collapsing isothermal and spherical dense core, the mass-infall rate ($\dot{M}_\mathrm{inf}$) is calculated to be $\sim$2$\times$10$^{-6}$ $M_\odot$ yr$^{-1}$  \citep{1977Shu}.
In IRAS~16544, in addition to the streamers, the extended gas around the protostar observed with ALMA is also infalling (see Section~\ref{subsec:model}). With the dense core mass within a radius of 4000 au measured with JCMT and the free-fall velocity at this radius calculated from the central protostellar mass, the mass infalling rate in the protostellar envelope is estimated to be %$\dot{M}_\mathrm{inf}$ $=$ 
1 to 5$\times$10$^{-6}$~$M_\odot$~yr$^{-1}$ in IRAS~16544, which is similar to the conventional picture and those observed in other Class 0 protostars \citep[e.g.,][] {1997Ohashi,1998Momose, 2013Takakuwa, 2015Aso, 2019Yen}. 
This is one or two orders of magnitude higher than the estimated mass infalling rate of the streamers. 
%We then use the JCMT C$^{18}$O moment 0 map to calculate the mass-infall rate in the envelope with a radius of 4000 au which is the consistent with the radius of streamers in 3D and derive the resulting $\dot{M}_\mathrm{inf}$ $=$ 1.6$^{+1.5}_{-0.2}$ $\times$ 10$^{-6}$ $M_\odot$ yr$^{-1}$. Adopting the enclosed mass within the radius, the mass infall rate is $\dot{M}_\mathrm{inf}$ $=$ 2.2$^{+3.0}_{-0.4}$ $\times$ 10$^{-6}$ $M_\odot$ yr$^{-1}$.
%The main errors in these values are attributed to be the C$^{18}$O abundance and excitation temperature.
%$\dot{M}_{streamer}$ in IRAS 16544 are extremely small compared to the typical mass infall rate in the early stages. 
Therefore, the streamers in IRAS~16544 most likely have limited contribution to the mass infall onto the central protostar+disk system. 

The mass accretion rate ($\dot{M}_\mathrm{acc}$) onto the protostar in IRAS~16544 is estimated to be 3.0$\times$10$^{-7}$~$M_\odot$~yr$^{-1}$ from the luminosity of the protostar as:  
%To study how much influences streamers have on disk formation in this object, we compare mass-infall rate via streamers and disk-to-star accretion rate ($\dot{M}_\mathrm{acc}$). $\dot{M}_\mathrm{acc}$ can be calculated given the accretion luminosity ($L_\mathrm{acc}$) and stellar radius ($R_\mathrm{*}$), 
\begin{equation}
    \dot{M}_{acc} = \frac{L_{acc}R_*}{GM_*}, 
\end{equation}
where ${L}_\mathrm{acc}$ is the accretion luminosity, $R_*$ is the stellar radius, $M_*$ is the stellar mass, and $G$ is the gravitational constant. 
The bolometric luminosity ($L_\mathrm{bol}$) of IRAS~16544 is 0.89~$L_\odot$ \citep{2023Ohashi}, which is the sum of the stellar luminosity ($L_{*}$) and the accretion luminosity ($L_\mathrm{acc}$). 
Assuming that the accretion luminosity is 50$\%$ of the bolometric luminosity \citep{2008Dunham, 2008Antoniucci} and the stellar radius is 3~$R_\odot$ \citep{1988Stahler}. 
This estimation tells us that the mass infalling rate of the streamers is a factor of several lower than the mass accretion rate onto the protostar or is comparable to it if we consider the conservative upper limit of the mass infalling rate of the streamers (Section~\ref{subsubsec:dis_Degeneracy}).
%the streamers have projected lengths of $\sim$1300 au.

If the streamers are the main mass supply to the disk and fully accreted onto the disk, the total disk$+$streamer mass would be consumed in $\sim$3$\times$10$^4$~yr by the mass accretion from the disk onto the protostar, assuming that the mass accretion onto the protostar is constant.
%We also calculated the disk lifetime, considering only the gas within the disk and contribution from the streamers, under the assumption that the mass accretion rate is constant, resulting in $\sim$3$\times$10$^4$ yr.
Then the disk lifetime would be shorter than the typical time scale of the Class 0 and I stages \citep[e.g.,][]{2022Mercimek}.
%The mass accretion rate from the disk onto the protostar does not vary significiantly with time, the mass supply to the disk from the infalling streamers is insufficient to maintain the disk, and the disk mass may be fully consumed by the accretion onto the protostar before the protostar evolves to the Class II stage.
%If there are only streamers, the protostellar disk can be alive until $\sim$ 3-6$\times$10$^4$ yr, which means the disk will disappear before evolving to Class II stage. 
%On the other hand, if all gas in the dense core transported to the disk, its lifetime is extended to be 1.8$\times$10$^6$ yr, which is comparable to the typical YSO's disk lifetime of 2-3 $\times$10$^6$ yr \citep{2009Mamajek, 2011Williams}.
%The contribution of streamers in IRAS 16544 is limited to maintain the disk rather than the isotropically infalling gas.
Hence, the mass infall from the extended protostellar envelope onto the disk, in addition to the streamers, likely plays a more important role in the star formation process in IRAS~16544, unless the streamers are connected to a larger mass reservoir beyond the field of the view of our ALMA observations. 
%maintaining the disk, rather than the mass supply through streamers.

In other protostars exhibiting streamers, the mass infalling rates of the streamers are mostly on the order of 10$^{-6}$~$M_\odot$~yr$^{-1}$ \citep[e.g.,][]{2017Yen,2021Pineda,2022Thieme,2022Valdivia,2023Flores,2023Hsieh,2024Cacciapuoti,2024Taniguchi,2024Lin}. 
%\textcolor{blue}{cheong+2018}
%master thesis
This is comparable to the mass infalling rate in the classical model of collapsing dense cores. In these protostars, their streamers may play a more significant role in the star formation process, unlike the case of IRAS~16544. On the other hand, some other protostars have streamers with mass infalling rates on the order of 10$^{-7}$~$M_\odot$~yr$^{-1}$ \citep{2014Yen,2022Thieme, 2022Murillo, 2024Codella}. 
%The difference of the mass infalling rates of streamers between the other sources and IRAS~16544 may come from the protostellar mass. This source has relatively smaller stellar mass compared to other sources, and thus the mass infalling rate of the streamers is smaller.
%In these protostars, the mass infalling rates of the streamers are higher than IRAS~16544 by a factor of several to two orders of magnitude. Thus, 
More statistical studies on streamers compared to global infall in protostellar envelopes are needed to understand their importance in star formation. 

Even though in IRAS~16544 the mass infalling rate of the streamers is low, on the order of 10$^{-8}$~$M_\odot$~yr$^{-1}$, the streamers may still be able to cause accretion shocks on the disk, as demonstrated in numerical simulations \citep{2015Lesur}. 
Indeed, typical shock tracers, such as SO and SiO \citep[e.g.,][]{2015Aota, 2017Miura, 2021vanGelder, 2023Yamato, 2023Flores, 2023van'tHoff}, have been detected in the NE streamer \citep{2023Kido}, which could be due to sublimation/sputtering of dust grains caused by the mutual interactions between the NE streamer and protostellar disk. Observations resolving the shock regions in molecular lines are essential to reveal the physical and chemical impact of these streamers on the disk. 

\subsection{Formation mechanisms of streamers}\label{subsubsec:dis_origin_of_streamers}
Theoretically, streamer-like structures can form via several mechanisms: (1) collapse of turbulent and magnetized dense cores \citep[e.g.,][]{2023Tu, 2015Seifried}, (2) interactions with stellar flyby \citep[e.g.,][]{2023Cuello, 2023Smallwood}, (3) encounter with cloudlet components \cite[e.g.,][]{2019Dullemond, 2024Hanawa}, and (4) collision between dense cores or between dense cores and filaments \citep{2024Yano, 2024Nakamura}.

A flyby is a perturber encountering a central star in a parabolic or hyperbolic orbit. If the central star is harboring a disk, 
%when a perterbur encounters the star at periastron, 
their mutual interaction can lead to the formation of spirals and a bridge connecting the two stars, and such structures may appear like streamers \citep{2023Cuello,2024Smallwood}.  
%disk truncation, and change of the disk eccentricity \citep{2023Cuello,2024Smallwood}. 
%\citep{1994Ostriker, 2003Pfalzner,2010Shen, 2010Thies, 2023Smallwood, 2024Smallwood}
IRAS~16544 is embedded in the globule CB~68 located at the edge of the filament in the Ophiuchus North region. It is the only compact source detected within a radius of 100$\arcsec$ ($\sim$ 15,000~au) from the globule \citep{1995Wang, 1999Huard}.
%The globule has only one IRAS point source within a radius of 100$\arcsec$ ($\sim$ 15000 au) \citep{1995Wang, 1999Huard}. 
The closest objects to IRAS~16544 on the plane-of-sky are located 15$\arcmin$ ($\sim$1.3$\times$10$^5$~au) away, which are young stellar objects \citep{2000Tachihara, 2012Hatchell}.
%, one IRAC source in the southeastern region \citep{2012Hatchell} and one IRAS source in the northwestern region  \citep{2000Tachihara, 2012Hatchell}. 
%The q2 core, which is harboring an IRAS source is located 15$\arcmin$ ($\sim$1.3$\times$10$^5$ au) to the northwest \citep{2000Tachihara, 2012Hatchell}. 
%The bridge-like component connecting between IRAS 16544 and q2 core is detected by Herschel continuum maps at 100, 160, 250, 350, and 500 $\mu$m and Digitized Sky Survey (DSS2) optical red image \citep{2013Launhardt}. 
%There is the possibility that the streamers detected by ALMA are part of the bridge component. It might be useful to observe a wide spatial scale in several wavelength to measure the mass and gas kinematics of the bridge component. 
Therefore, it is unlikely that a flyby encounter has occurred in IRAS~16544.
%If the formation of a bridge component due to a stellar flyby is occurred within 1000 au by definition, then streamers are observed in this source are unlikely to be explained by a flyby.

%The C$^{18}$O moment 0 map observed by JCMT shows only a single peak in the dense core, and there are no structures similar to a cloudlet.
%\textbf{We note that the NE streamer was also seen in the CCH emission, which may hint at chemically fresh material, the CCH distribution is possibly affected by the outflows \citep{2022Imai}.}
%\textbf{These are opposite} to the case of IRAS~03292$+$3039, where the streamer is connected with a cloudlet with a size of 8000~au and 0.1~pc ($\sim$2$\times$10$^4$~au) away from the protostar \textbf{and is mainly detected with tracers of chemically fresh material} \citep{2024Taniguchi}.
Toward IRAS~03292$+$3039, a streamer component seen in molecular tracers of chemically-fresh material is connected with a cloudlet $\sim$8000~au in size, which is $\sim$0.1~pc away from the protostar \citep{2024Taniguchi}. This is an example of streamers produced by the cloudlet component. While the streamers in IRAS~16544 are also traced by the CCH emission, a prototypical tracer of chemically-young molecular gas \citep{2022Imai}, the JCMT C$^{18}$O moment 0 map shows only a single peak in the dense core without any cloudlet component.
%where \cite{2024Taniguchi} detected a star-less like component with the size of 0.04 pc at $\sim$20,500 au northeast toward the protostar. They suggested that this component might be the reservoir for the streamer since it is located in the same direction as the streamer and show a velocity gradient connecting it to protostar.
%In the case of IRAS 16544, two or more peaks are not detected within the radius of 100$\arcsec$ by JCMT and CSO, however, observing on a larger spatial scale could potentially reveal a reservoir-like component at the tips of streamers.
%In addition, the gas free-fall time of 5-13$\times$10$^4$ yr
Therefore, the streamers in IRAS~16544 are unlikely to be caused by the accretion from passing cloudlets. 
%The accretion from the reservoir seems unlikely to form streamers in IRAS 16544 that we observe. 
%However, it is important to note that the resolution of our JCMT observations is limited to 21$\arcsec$ (or 3200 au) and any cloudlets with a size of a few thousand au may be smoothed out. 
However, it is important to note that the resolution of our JCMT observations, 21$\arcsec$ (or $\sim$3200~au), could smooth out 1000~au sized cloudlets, if any.
Future mosaic observations at high resolution over the wider area will help us to reveal the density distribution in the globule CB~68 in more detail.
%The possibility of mass reservoir from the cloudlet is unlikely to be clarified by the time wide-field mosaic observations are conducted.
%Furthermore, since the streamers are detectable only with ALMA data, it is considered that the streamers in this object originate from accretion flows from the envelope to the disk.

When a protostellar dense core collides with another dense core or a filament, a shocked-compressed layer can form. This layer may gradually transform into a spiral structure connecting to the disk by the stellar gravity, and eventually appear as a streamer \citep{2024Yano, 2024Nakamura}. %Such collision may form shocks and shocked gas tracers are expected to be likely abundant in the region.
%when two isothermal spherical cores or the core and a filament collide, the compressed region at the collision interface is gradually connected with a rotational disk and gradually similar to the shape of streamer. Such region is heated up to $<$ 100 K by the (supersonic shock?) collision shock and thus the shocked gas tracers, SO, SiO and CH$_3$OH, are expected to be likely abundant.
%Compared to \cite{2024Yano}, there is a possibility that these emissions are not caused by accretion but by the collision. However, even if shocked gas tracers become abundant along the compressed region between cores, it remains unclear how their abundance and spatial distribution evolve by the time arm-like structures are connected to the disk. In particular, whether the shocked gas tracers eventually fall near the disk cannot be determined without solving the chemical network in their model.
In the JCMT C$^{18}$O data, only a single velocity component is observed (Figure~\ref{fig_JCMTline}), and the non-thermal linewidth is small ($\Delta$$V_\mathrm{NT}$ $\leq$ 1.0~km~s$^{-1}$). Although the shocked gas tracers have been detected in IRAS~16544 \citep{2023Kido}, they are in the vicinity of the central disk.
Hence, there is no sign of collision between the dense core of IRAS~16544 and another component.
%For CFC model, \cite{2024Nakamura} carried out to observe IRAS 03292 with Nobeyama Radio Observatory 45m telescope in CCS ($J_N$ $=$ 4$_3$--3$_2$) emission and detected the gas flow connecting the star-less core to protostellar core. In the PV diagram along these components, it displayed a velocity structure resembling a Hubble outflow \citep{2007Arce}, where higher velocity components are detected farther from the protostrar, rather than a velocity gradient indicative of gravitational infall toward the star. Therefore, they suggested that the elongated structure to the starless core is formed by the collision between dense core and filament because the velocity gradient should be consistent with outflow if so and stellar gravity is not dominant. In our target source, we do not see this feature in PV diagrams and thus it is unlikely that streamers are not formed by CFC scenario.  

Simulations of the collapse of magnetized dense cores, both with and without initial turbulence, often show several accretion flows connecting to the disk, which are similar to streamers in observations 
\citep{2017Kuffmeier, 2019Lam, 2023Tu, 2024Lebreuilly}.
%The cloud collapse in the magnetized environment with and without initial turbulence shows several accretion flows connecting to the disk, which is similar to the streamer \citep{2023Tu}. 
%Without turbulence, the magnetic flux is released from the material accreated to the central star and its pressure pushes ambient gas and forms low-density cavity. The material lines along the edge of the cavity and looks like streamers. These structures are concentrated on the midplane. 
Without turbulence, the high magnetic pressure pushes ambient gas away and forms low-density cavities when the magnetic flux is released from the material accreted onto the protostar. Streamer-like accretion flows then appear along these cavities. These accretion flows are most concentrated in the equatorial plane of the disk around the central star. 
With turbulence, the pseudo disk can be warped or fragmented and appears as accretion flows. These flows often extend vertically away from the equatorial plane of the disk. %The formation mechanism of streamers differs depending on whether turbulence is present or not. 

Our analysis shows that IRAS~16544 has subsonic to transonic turbulence and a mass-to-flux ratio of 2--6 in the dense core.
These levels of turbulence and magnetic field strength are comparable to those in the simulations of \cite{2023Tu}. 
In addition, the streamers in IRAS~16544 likely extend away from the midplane. 
Thus, the streamers in IRAS~16544 could be similar to those formed in the simulations of magnetized and turbulent dense cores.
%have height in the z-axis direction, not on the midplane. These signatures are consistent with the accretion model including the turbulence.
%For the ambipolar diffusion, differences in the magnetic diffusivity show a clear trend in the formation of streamers and disk size.
%In the lower ambipolar diffusivity model, envelope is more fragmented and heterogeneous and the size of the formed disk is smaller than the higher magnetic diffusivity environment. 
%The dense core contains a compact gas disk with a radius of 55 au and three streamers with the widths of $<$ 6$\farcs$2 (MRS).
%If streamers are formed in such a region, the dense core of IRAS 16544 might be in a relatively magnetic diffusive environment. 
%If the ionization rate within the envelope can be measured in future studies, it will become possible to discuss the diffusivity of the magnetic field within the dense core.
%Whether there is a correlation between disk size and streamer structure due to the ambipolar diffusivity could potentially be clarified by measuring it in the envelope. 
However, these simulations show that the streamers are the main channels of mass accretion, and the density is enhanced by a factor of $\sim$30 in the streamers compared to the initial mean density of the dense core \citep{2023Tu}. In contrast, the streamers in IRAS~16544 have limited contribution to mass accretion toward the central protostar$+$disk system, and their density is only a factor of a few higher than the mean density of the dense core observed with JCMT. Thus, the streamers observed in IRAS~16544 are not as prominent as those seen in the simulations. While we observe streamers in IRAS~16544, their underlying formation mechanisms remains unclear.
Further comparison of the density and velocity structures of the streamers in the theoretical simulations and observations is required to clarify the difference. 
%For the surface density, the contrast in streamers and dense core of their model and observations is $\sim$28 and $\sim$4.7, respectively. The effect of missing flux in streamers is minimal since the flux does not change when 7-m array data is added. In the simulation, mass supply via streamers is the main contributor to accumulating mass in the disk, while in the case of IRAS 16544, the contrast is one order of magnitude smaller, and their contribution is also reduced.

Thus, the origin of the streamers in IRAS~16544 is not adequately explained with the formation mechanisms mentioned above. We note that on an even larger scale, there is a bridge component connecting the dense core of IRAS~16544 and the C$^{18}$O core q2 which has the characteristic of a YSO, located 15$\arcmin$ ($\sim$1.3$\times$10$^5$~au) to the northwest \citep{2000Tachihara, 2012Hatchell}.
This is detected by the Herschel continuum maps at 100, 160, 250, 350, and 500~$\mu$m and Digitized Sky Survey (DSS2) optical red image \citep{2013Launhardt}. 
%(May mention the results of Pineda and their group showing addition mass supply to the dense core by large-scale streamers.)
Streamers connecting to central protostellar systems and extending beyond associated dense cores have been observed in other protostars \citep{2020Pineda}, suggesting that the material may be funneled from outside the dense core. %found a streamer on 10,5000 au scale and proposed that material are funneled from outside the dense core via large-scale streamers.
In IRAS~16544, all the prominent streamers are observed on the northern side in the dense core, where there is a large-scale bridge component. 
However, the relation between the streamers and the large scale bridge is not clear.
To explore the potential connection between the streamers in IRAS~16544 and this bridge, high resolution observations covering a broader area are essential. These observations can clarify if the streamers in IRAS~16544 are part of a larger structure funneling mass transfer from the bridge to the protostar.

\section{Summary}  \label{sec:summary}
We have analyzed the C$^{18}$O emission observed with JCMT and ALMA to investigate the kinematics and physical properties of the dense core, protostellar envelope, and streamers in IRAS~16544. The main results of our analysis are summarized as follows:
\begin{itemize}
    \item [1.]
    The mass of the dense core derived from the JCMT observation is estimated to be $\sim$0.52-0.60~$M_\odot$. %, consistent with estimation from dust continuum emissions reported in previous papers.
    The Mach number ranges from 0.8 to 1.1, suggesting that the dense core has a subsonic to transonic level of turbulence.
    With the magnetic angular dispersion reported in \cite{2024Yen}, the magnetic field strength is estimated to be 23--76~$\mu$G, corresponding to a dimensionless mass-to-flux ratio of 2--6, making the core supercritical. 
    \item[2.]
    Elongated structures with projected lengths of $\sim$1300~au are detected on the northern side of the disk by the ALMA observations. 
    Their spatial and velocity structures can be explained with the kinematical model of free-falling streamers with a conserved angular momentum and zero total energy. %the accretion model in which material flows from the outskirts of the dense core to the disk while conserving angular momentum, and they are identified as infalling streamers. 
    Their lengths in 3D space are estimated to be 2800--6100~au based on the infall model. The mass and mass infalling rate of these streamers are estimated to be (1--4)$\times$10$^{-3}$~$M_\odot$ and (1--5)$\times$10$^{-8}$~$M_\odot$~yr$^{-1}$, respectively. The impact of missing flux is limited, based on a comparison with the JCMT$+$ALMA combined map. In addition to the streamers, there is a more extended protostellar envelope observed in the JCMT$+$ALMA combined map, and its mass infalling rate is estimated to be (1--5)$\times$10$^{-6}$~$M_\odot$~yr$^{-1}$ calculated from the JCMT observation. 
    Compared to the protostellar envelope, the contribution of mass supply to the disk via the streamers is relatively small.
    \item[3.]
    The streamers in IRAS~16544 are unlikely formed due to a flyby, accretion of a cloudlet, or core-core/filament collision. 
    The dense core of IRAS~16544 is turbulent and magnetized.
    Numerical simulations of collapsing dense cores with similar conditions often form streamers \citep[e.g.,][]{2023Tu, 2015Seifried}. In these simulations, the mass contribution from the streamers is dominant compared to the envelope, and they are the main channels of mass accretion. In contrast, the streamers in IRAS~16544 have a total mass of 7.5$\times$10$^{-3}$~$M_\odot$, only accounting for $2\%$ of the dense core mass. This is different from those numerical simulations.    
    The origin of the streamers in IRAS~16544 is yet to be elucidated by future observational and theoritical studies.
    %The streamers only play a minor role in mass accretion onto the disk in IRAS~16544.
    %\item[4.]
    %We found that the free-fall solutions with the conserved angular momentum are not unique to reproduce the observed streamer features. Due to the projection effect, degeneracy of the model solutions is present for each of the observed streamers. On the other hand, such a degeneracy does not alter our conclusion that the observed streamers in IRAS 16544 are minimal as compared to the total budget of the mass accretion originated from the more isotropic protostellar envelope.
\end{itemize}

\section*{Acknowledgments}
We are grateful to the anonymous referee for the helpful and constructive comments.
We would like to thank all the ALMA staff supporting this work. 
This paper makes use of the following ALMA data: ADS/JAO.ALMA \#2018.1.01205.L, 2019.1.00261.L, and  2019.A.00034.S.
ALMA is a partnership of ESO (representing its member states), NSF (USA), and NINS (Japan), together with NRC (Canada), MOST and ASIAA (Taiwan), and KASI (Republic of Korea), in cooperation with the Republic of Chile. 
The Joint ALMA Observatory is operated by ESO, AUI/NRAO, and NAOJ. 
The National Radio Astronomy Observatory is a facility of the National Science Foundation operated under cooperative agreement by Associated Universities, Inc. 
%M. K. is supported by the Academia Sinica TIGP-X program (AS-CDA-111-M03 and NSTC 110-2628-M-001-003-MY3) and by JSPS KAKENHI grant Number 24KJ1834.
M.K. is supported by the TIGP-X Pilot Program, Academia Sinica, Taipei,
Taiwan and by JSPS KAKENHI grant Number 24KJ1834.
H.-W.Y. acknowledges support from the National Science and Technology Council (NSTC) in Taiwan through the grant NSTC 113-2112-M-001-035- and from the Academia Sinica Career Development Award (AS-CDA-111-M03).
S.T. is supported by JSPS KAKENHI grant Nos. 21H00048 and 21H04495, and by NAOJ ALMA Scientific Research grant No. 2022-20A.
N.O., C. F., and M.N. acknowledge support from National Science and Technology Council (NSTC 113-2112-M-001-037) and the Academia Sinica Investigator Project Grant (AS-IV-114-M02).
Y.A. acknowledges support by NAOJ ALMA Scientific Research Grant code 2019-13B, Grant-in-Aid for Scientific Research (S) 18H05222, and Grant-in-Aid for Transformative Research Areas (A) 20H05844 and 20H05847.
I. H. acknowledges support the funding from the European Research Council (ERC) under the European Union’s Horizon 2020 research and innovation programme (Grant agreement No. 101098309 - PEBBLES)
P.M.K. acknowledges support from NSTC 108-2112- M-001-012, NSTC 109-2112-M-001-022 and NSTC 110-2112-M-001-057.
W.K. was supported by the National Research Foundation of Korea (NRF) grant funded by the Korea government (MSIT) (RS-2024-00342488).
J.-E.L. was supported by the National Research Foundation of Korea (NRF) grant funded by the Korea government (MSIT) (grant numbers 2021R1A2C1011718 and RS-2024-00416859).
Z.-Y.L. is supported in part by NASA NSSC20K0533, NSF AST-1910106 and Virginia Institute for Theoretical Astronomy.
L.W.L. acknowledges support from NSF AST-2108794.
R.S acknowledge support from the Independent Research Fund Denmark (grant No. 0135-00123B). 
K. T. is supported by JSPS KAKENHI grant Nos. 21H04495, 21H04487, and 22KK0043.
J.P.W. acknowledges support from NSF AST-2107841.
%\textcolor{blue}{This study was supported by JSPS KAKENHI grants (24KJ1834, 21H00048, 21H04495).?}

\software{CASA \citep{Mcmullin2007}, matplotlib \citep{Hunter2007}, PVextractor \citep{2016Ginsburg}, astropy \citep{2022Astropy}}
\facility{JCMT, ALMA}

\appendix
\renewcommand{\thefigure}{A\arabic{figure}}
\renewcommand{\thetable}{A\arabic{table}}
\setcounter{figure}{0}
\setcounter{table}{0}
%\twocolumn

\section{Alternative models of infalling trajectories}
\label{appndix:Alternative model}
\begin{figure}[h]
\centering
\includegraphics[width=1.0\textwidth, angle=0]{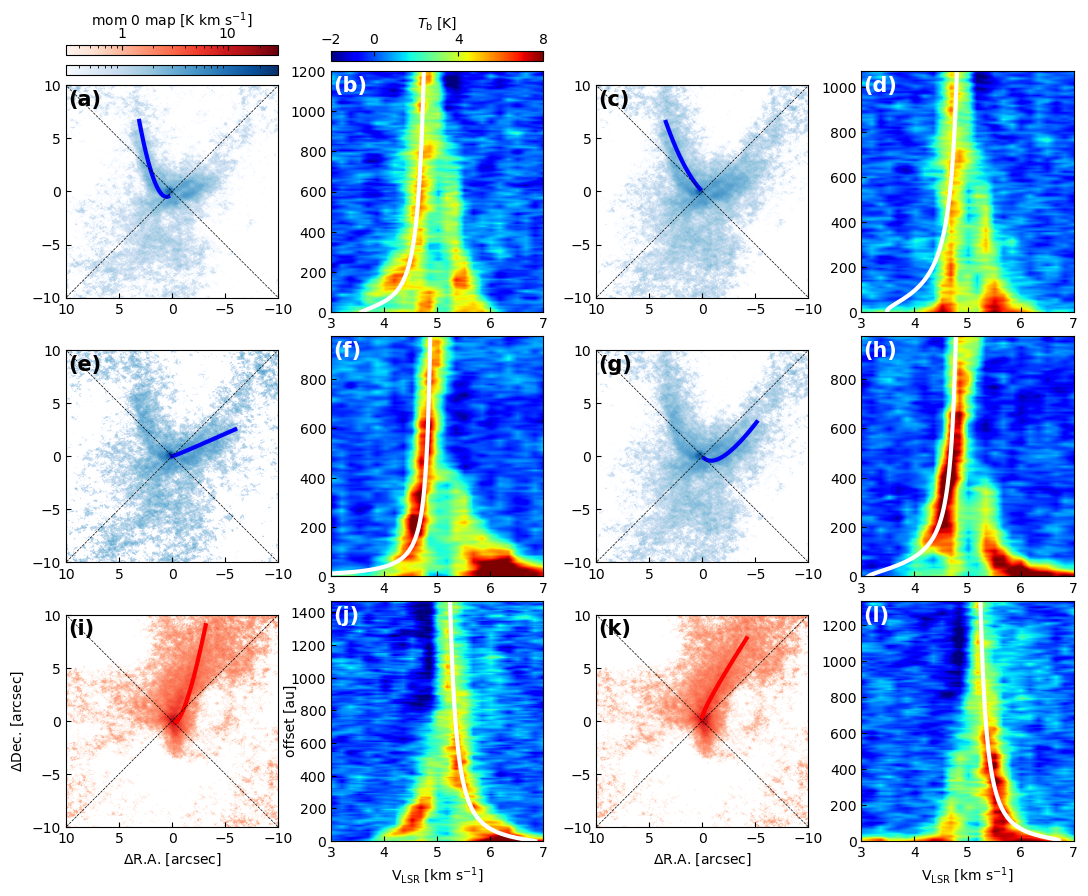}\\
\caption{Moment 0 maps of the C$^{18}$O emission integrated over the velocity ranges of each streamer (first and third columns) and PV diagrams cut along the solid lines in each left panel (second and forth columns). 
%First column is same as Figure~\ref{fig_almamom0}, but for integrating the velocity ranges od each streamer. Second column is same as Figure~\ref{fig_streamerpv} and 
These panels present the spatial and velocity structures of the different model infalling trajectories that can explain the observations. The first and second columns show our fiducial models adopted in the main text, whereas the third and forth columns show other possible models. The model parameters are summarized in Table \ref{table:different model}.}
\label{fig_different_model}
\end{figure}

Figure~\ref{fig_different_model} compares the spatial and velocity structures of the adopted and different model parameters that can explain the observed moment 0 maps and PV diagrams.
%presents the spatial and velocity structures of infalling trajectories computed with different model parameters in comparison with the observed moment 0 maps and PV diagrams. 
The parameters of each streamer are listed in Table~\ref{table:different model}. The C$^{18}$O moment 0 maps are generated by integrating the emission over the velocity ranges of each streamer. With the parameters different from those adopted in Section~\ref{subsec:streamers by ALMA}, the model trajectories can also explain the observations. 

\begin{table}[h]
    \centering
    \caption{Adopted parameters in the UCM model}
    \begin{tabular}{cccccccccccc}
    \hline\hline
    \colhead{} & \colhead{} & \colhead{$i_\mathrm{s}$}& \colhead{$\theta_\mathrm{s}$}& \colhead{$r_\mathrm{d}$}& \colhead{$\theta_0$}& \colhead{$\phi_0$} & \colhead{$r$} & \colhead{$\Delta V$}&\colhead{$M$}& \colhead{$t$}& \colhead{$\dot{M}_\mathrm{inf}$}\\
    
    \colhead{} & \colhead{} & \colhead{($\degr$)}& \colhead{($\degr$)}& \colhead{(au)}& \colhead{($\degr$)}& \colhead{($\degr$)} & \colhead{(au)} & \colhead{(km s$^{-1}$)}& \colhead{($M_\odot$)}& \colhead{(yr)}& \colhead{(M$_\odot$ yr$^{-1}$)}\\
    %\colhead{$r$}\\
    \hline
    NE & fiducial model &  73 & 60 & 100 & 90& 64 & 4790 & 3.6--4.7 &1.3$^{+1.4}_{-0.2}$$\times$10$^{-3}$ & 1.0$\times$10$^5$ & 1.3$^{+1.4}_{-0.2}$$\times$10$^{-8}$\\
     & different model parameter& 107 & $-$45 & 55 & 70& 73 &16830 &3.5--4.8&1.9$^{+1.8}_{-0.3}$$\times$10$^{-3}$ & 6.6$\times$10$^5$ & 2.8$^{+2.8}_{-0.4}$$\times$10$^{-9}$\\\hline
    NW & fiducial model & 107& $-$45 &55 & 20& $-$10 & 2830 &$-$0.1--4.9  &2.2$^{+2.1}_{-0.4}$$\times$10$^{-3}$ & 4.5$\times$10$^4$ & 4.8$^{+4.7}_{-0.7}$$\times$10$^{-8}$\\
     & different model parameter & 150& 120 &100& 40& 68 &7520 & 3.1--4.8&2.0$^{+1.9}_{-0.3}$$\times$10$^{-3}$&2.0$\times$10$^5$&10$^{+10}_{-1.4}$$\times$10$^{-9}$\\\hline
    NNW & fiducial model &107 & $-$45& 55& 70& $-$120& 6130 & 5.2--6.8 &4.0$^{+4.0}_{-0.6}$$\times$10$^{-3}$ & 1.4$\times$10$^5$ & 2.8$^{+2.7}_{-0.4}$$\times$10$^{-8}$\\
     & different model parameter & 30& $-$30 & 100& 20& 75 & 2590 & 5.2--6.7& 4.6$^{+4.4}_{-0.7}$$\times$10$^{-3}$& 4.0$\times$10$^4$ &11$^{+12}_{-1.2}$$\times$10$^{-8}$\\\hline
     \end{tabular}
     
\tablecomments{$i_\mathrm{s}$ is defined as the angle between the rotational axis and the LOS, from 0$\degr$ to 180$\degr$. $\theta_\mathrm{s}$ is defined as the angle between the north and the rotational axis and increasing counterclockwise, from 0$\degr$ to 360$\degr$. $\theta_\mathrm{0}$ $=$ 0$\degr$ is toward the north and it changes from 0$\degr$ to 180$\degr$. $\phi_0$ $=$ 0$\degr$ indicates the west on the plane-of-sky when $\theta_\mathrm{s}$ $=$ 0$\degr$ and increases counterclockwise, from 0$\degr$ to 360$\degr$.}
\label{table:different model}
\end{table}
%First column in Fig.~\ref{fig_different_model} are the moment 0 maps integrating the C$^{18}$O emission over the velocity ranges of each streamer listed in Table~\ref{table:different model}. Second column in Fig.~\ref{fig_different_model} are same as Fig.~ \ref{fig_streamerpv}, whereas third and forth columns in Fig.~\ref{fig_different_model} shows the trajectories and PV diagrams adopting different model parameters. 
%For example, a comparison between the UCM model with $i_s$ $=$ 120$\degr$, $\theta_s$ $=$ 150$\degr$, $r_d$ $=$ 100 au, $\theta_0$ $=$ 40$\degr$, and $\phi_0$ $=$ 78$\degr$ and observation is shown in Figure \ref{fig_different_model}(g, h).
%Fig.~\ref{fig_different_model} (g, h) shows other UCM model which can represent NW streamer with the parameters of $i_\mathrm{s}$ $=$ 150$\degr$, $\theta_\mathrm{s}$ $=$ 120$\degr$, $r_\mathrm{d}$ $=$ 100 au, $\theta_0$ $=$ 40$\degr$, and $\phi_0$ $=$ 68$\degr$. 

\section{Channel maps of the 12-m array data}
\renewcommand{\thefigure}{B\arabic{figure}}
\renewcommand{\thetable}{B\arabic{table}}
\setcounter{figure}{0}
\setcounter{table}{0}
\label{appndix:channelmap}
\begin{figure}[t]
\centering
\includegraphics[width=180mm, angle=0]{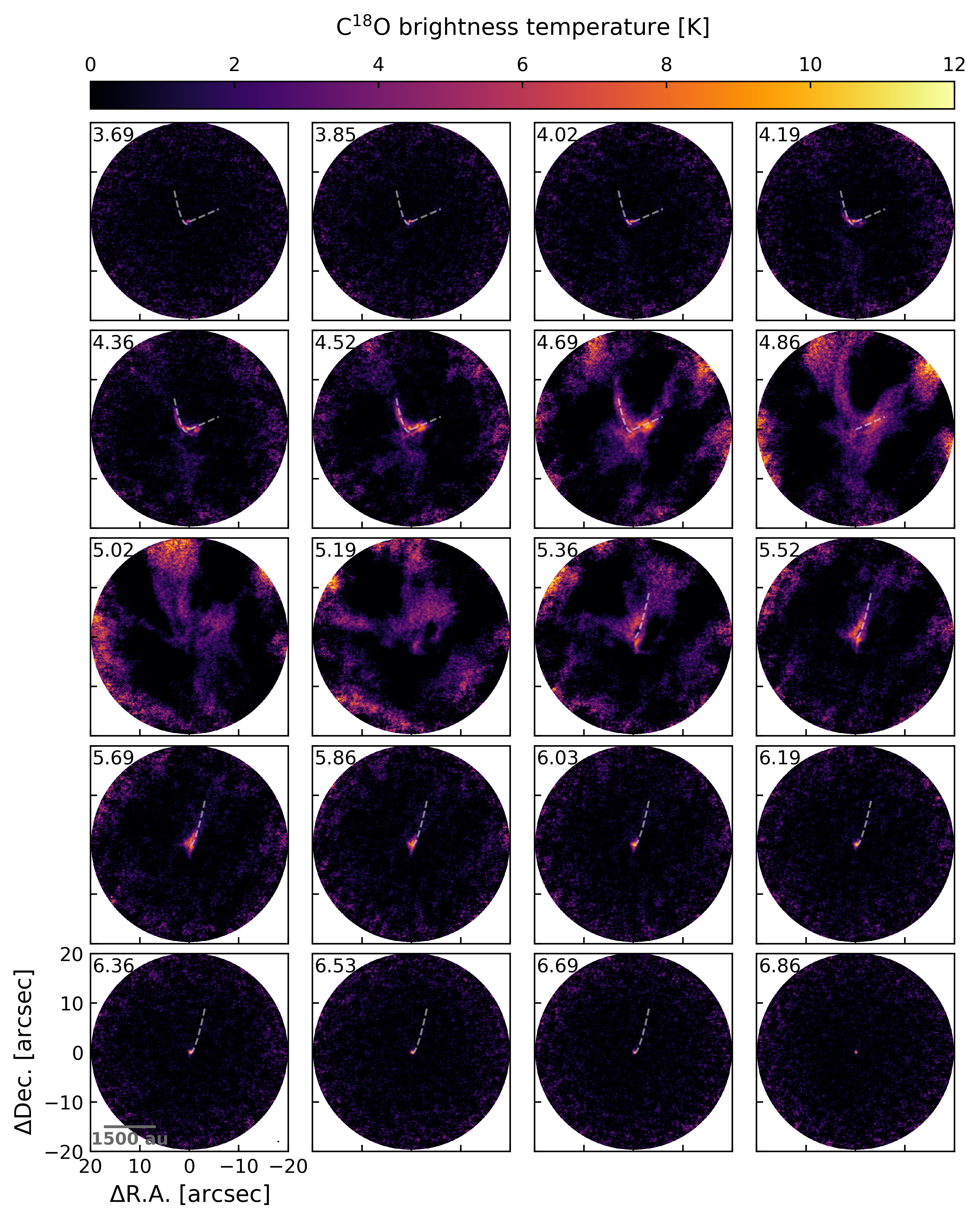}\\
\caption{Same as Figure~\ref{fig_channelmap}, but for the ALMA 12-m array data only without the 7-m array data. The beam size is 0$\farcs$30$\times$0$\farcs$22 (P.A. $=$ 77$\degr$) and the rms noise level is 1.35~mJy~beam$^{-1}$ ($=$ 0.52~K).}
\label{fig_channelmap_12only}
\end{figure}
Figure~\ref{fig_channelmap_12only} shows the channel maps of the C$^{18}$O emission generated from the ALMA 12-m array data without the 7-m array data.
The images are generated with the same procedure in Section~\ref{datareduction:ALMA} but with the auto-masking parameters of the sidelobe threshold of 1.0, noise threshold of 3.5, low noise threshold of 0.5, minimum beam fraction of 0.3, and negative threshold of 7.0. 
Strong emission appears at the edge of the field-of-view after primary beam correction. 
This is particularly prominent in the low-velocity range of 4.69--5.52~km~s$^{-1}$, and it is more significant than that in the 12-m and 7-m array combined maps (Figure~\ref{fig_channelmap}). 
Thus, there is a caution of potential artifacts, when discussing streamer-like structures using the 12-m array only data.

Without the 7-m array data, the mean missing flux at the velocities of $V_\mathrm{LSR}$ $=$ 4.0--4.9~km~s$^{-1}$, $V_\mathrm{sys}$, and $V_\mathrm{LSR}$ $=$ 5.2--6.0~km~s$^{-1}$ are estimated to be 74$\%$, 97$\%$, and 87$\%$, respectively.
Including the 7-m array data, the recoverable flux is only slightly improved compared to the 12-m array only data.

%Figure~\ref{fig_channelmap_feather} presents the velocity channel maps the C$^{18}$O emission generated with all combined data, including the JCMT and ALMA 7-m and 12-m array data. 
%shows the channel maps of the C$^{18}$O emission generated with all combined feather data. The flux around the systemic velocity ($V_\mathrm{sys}$ $=$ 4.86-5.52 km s$^{-1}$) can be recovered.

\bibliographystyle{aasjournal}
\bibliography{CB68_paper}

\begin{thebibliography}{}
\expandafter\ifx\csname natexlab\endcsname\relax\def\natexlab#1{#1}\fi
\providecommand{\url}[1]{\href{#1}{#1}}
\providecommand{\dodoi}[1]{doi:~\href{http://doi.org/#1}{\nolinkurl{#1}}}
\providecommand{\doeprint}[1]{\href{http://ascl.net/#1}{\nolinkurl{http://ascl.net/#1}}}
\providecommand{\doarXiv}[1]{\href{https://arxiv.org/abs/#1}{\nolinkurl{https://arxiv.org/abs/#1}}}

\bibitem[{{Andrews} {et~al.}(2018){Andrews}, {Huang}, {P{\'e}rez}, {Isella}, {Dullemond}, {Kurtovic}, {Guzm{\'a}n}, {Carpenter}, {Wilner}, {Zhang}, {Zhu}, {Birnstiel}, {Bai}, {Benisty}, {Hughes}, {{\"O}berg}, \& {Ricci}}]{2018Andrews}
{Andrews}, S.~M., {Huang}, J., {P{\'e}rez}, L.~M., {et~al.} 2018, \apjl, 869, L41, \dodoi{10.3847/2041-8213/aaf741}

\bibitem[{{Antoniucci} {et~al.}(2008){Antoniucci}, {Nisini}, {Giannini}, \& {Lorenzetti}}]{2008Antoniucci}
{Antoniucci}, S., {Nisini}, B., {Giannini}, T., \& {Lorenzetti}, D. 2008, \aap, 479, 503, \dodoi{10.1051/0004-6361:20077468}

\bibitem[{{Aota} {et~al.}(2015){Aota}, {Inoue}, \& {Aikawa}}]{2015Aota}
{Aota}, T., {Inoue}, T., \& {Aikawa}, Y. 2015, \apj, 799, 141, \dodoi{10.1088/0004-637X/799/2/141}

\bibitem[{{Arce} {et~al.}(2007){Arce}, {Shepherd}, {Gueth}, {Lee}, {Bachiller}, {Rosen}, \& {Beuther}}]{2007Arce}
{Arce}, H.~G., {Shepherd}, D., {Gueth}, F., {et~al.} 2007, in Protostars and Planets V, ed. B.~{Reipurth}, D.~{Jewitt}, \& K.~{Keil}, 245, \dodoi{10.48550/arXiv.astro-ph/0603071}

\bibitem[{{Aso} {et~al.}(2015){Aso}, {Ohashi}, {Saigo}, {Koyamatsu}, {Aikawa}, {Hayashi}, {Machida}, {Saito}, {Takakuwa}, {Tomida}, {Tomisaka}, \& {Yen}}]{2015Aso}
{Aso}, Y., {Ohashi}, N., {Saigo}, K., {et~al.} 2015, \apj, 812, 27, \dodoi{10.1088/0004-637X/812/1/27}

\bibitem[{{Aso} {et~al.}(2023){Aso}, {Kwon}, {Ohashi}, {J{\o}rgensen}, {Tobin}, {Aikawa}, {de Gregorio-Monsalvo}, {Han}, {Kido}, {Koch}, {Lai}, {Lee}, {Lee}, {Li}, {Lin}, {Looney}, {Narayanan}, {Phuong}, {Sai}, {Saigo}, {Santamar{\'\i}a-Miranda}, {Sharma}, {Takakuwa}, {Thieme}, {Tomida}, {Williams}, \& {Yen}}]{2023Aso}
{Aso}, Y., {Kwon}, W., {Ohashi}, N., {et~al.} 2023, \apj, 954, 101, \dodoi{10.3847/1538-4357/ace624}

\bibitem[{{Astropy Collaboration} {et~al.}(2022){Astropy Collaboration}, {Price-Whelan}, {Lim}, {Earl}, {Starkman}, {Bradley}, {Shupe}, {Patil}, {Corrales}, {Brasseur}, {N{\"o}the}, {Donath}, {Tollerud}, {Morris}, {Ginsburg}, {Vaher}, {Weaver}, {Tocknell}, {Jamieson}, {van Kerkwijk}, {Robitaille}, {Merry}, {Bachetti}, {G{\"u}nther}, {Aldcroft}, {Alvarado-Montes}, {Archibald}, {B{\'o}di}, {Bapat}, {Barentsen}, {Baz{\'a}n}, {Biswas}, {Boquien}, {Burke}, {Cara}, {Cara}, {Conroy}, {Conseil}, {Craig}, {Cross}, {Cruz}, {D'Eugenio}, {Dencheva}, {Devillepoix}, {Dietrich}, {Eigenbrot}, {Erben}, {Ferreira}, {Foreman-Mackey}, {Fox}, {Freij}, {Garg}, {Geda}, {Glattly}, {Gondhalekar}, {Gordon}, {Grant}, {Greenfield}, {Groener}, {Guest}, {Gurovich}, {Handberg}, {Hart}, {Hatfield-Dodds}, {Homeier}, {Hosseinzadeh}, {Jenness}, {Jones}, {Joseph}, {Kalmbach}, {Karamehmetoglu}, {Ka{\l}uszy{\'n}ski}, {Kelley}, {Kern}, {Kerzendorf}, {Koch}, {Kulumani}, {Lee}, {Ly}, {Ma}, {MacBride}, {Maljaars}, {Muna}, {Murphy}, {Norman},
  {O'Steen}, {Oman}, {Pacifici}, {Pascual}, {Pascual-Granado}, {Patil}, {Perren}, {Pickering}, {Rastogi}, {Roulston}, {Ryan}, {Rykoff}, {Sabater}, {Sakurikar}, {Salgado}, {Sanghi}, {Saunders}, {Savchenko}, {Schwardt}, {Seifert-Eckert}, {Shih}, {Jain}, {Shukla}, {Sick}, {Simpson}, {Singanamalla}, {Singer}, {Singhal}, {Sinha}, {Sip{\H{o}}cz}, {Spitler}, {Stansby}, {Streicher}, {{\v{S}}umak}, {Swinbank}, {Taranu}, {Tewary}, {Tremblay}, {Val-Borro}, {Van Kooten}, {Vasovi{\'c}}, {Verma}, {de Miranda Cardoso}, {Williams}, {Wilson}, {Winkel}, {Wood-Vasey}, {Xue}, {Yoachim}, {Zhang}, {Zonca}, \& {Astropy Project Contributors}}]{2022Astropy}
{Astropy Collaboration}, {Price-Whelan}, A.~M., {Lim}, P.~L., {et~al.} 2022, \apj, 935, 167, \dodoi{10.3847/1538-4357/ac7c74}

\bibitem[{{Bertrang} {et~al.}(2014){Bertrang}, {Wolf}, \& {Das}}]{2014Bertrang}
{Bertrang}, G., {Wolf}, S., \& {Das}, H.~S. 2014, \aap, 565, A94, \dodoi{10.1051/0004-6361/201323091}

\bibitem[{{Cacciapuoti} {et~al.}(2024){Cacciapuoti}, {Macias}, {Gupta}, {Testi}, {Miotello}, {Espaillat}, {K{\"u}ffmeier}, {van Terwisga}, {Tobin}, {Grant}, {Manara}, {Segura-Cox}, {Wendeborn}, {Klessen}, {Maury}, {Lebreuilly}, {Hennebelle}, \& {Molinari}}]{2024Cacciapuoti}
{Cacciapuoti}, L., {Macias}, E., {Gupta}, A., {et~al.} 2024, \aap, 682, A61, \dodoi{10.1051/0004-6361/202347486}

\bibitem[{{CASA Team} {et~al.}(2022){CASA Team}, {Bean}, {Bhatnagar}, {Castro}, {Donovan Meyer}, {Emonts}, {Garcia}, {Garwood}, {Golap}, {Gonzalez Villalba}, {Harris}, {Hayashi}, {Hoskins}, {Hsieh}, {Jagannathan}, {Kawasaki}, {Keimpema}, {Kettenis}, {Lopez}, {Marvil}, {Masters}, {McNichols}, {Mehringer}, {Miel}, {Moellenbrock}, {Montesino}, {Nakazato}, {Ott}, {Petry}, {Pokorny}, {Raba}, {Rau}, {Schiebel}, {Schweighart}, {Sekhar}, {Shimada}, {Small}, {Steeb}, {Sugimoto}, {Suoranta}, {Tsutsumi}, {van Bemmel}, {Verkouter}, {Wells}, {Xiong}, {Szomoru}, {Griffith}, {Glendenning}, \& {Kern}}]{2022CASATeam}
{CASA Team}, {Bean}, B., {Bhatnagar}, S., {et~al.} 2022, \pasp, 134, 114501, \dodoi{10.1088/1538-3873/ac9642}

\bibitem[{{Cassen} \& {Moosman}(1981)}]{1981Cassen}
{Cassen}, P., \& {Moosman}, A. 1981, \icarus, 48, 353, \dodoi{10.1016/0019-1035(81)90051-8}

\bibitem[{{Chandrasekhar} \& {Fermi}(1953)}]{1953Chandrasekhar}
{Chandrasekhar}, S., \& {Fermi}, E. 1953, \apj, 118, 113, \dodoi{10.1086/145731}

\bibitem[{{Codella} {et~al.}(2024){Codella}, {Podio}, {De Simone}, {Ceccarelli}, {Ohashi}, {Chandler}, {Sakai}, {Pineda}, {Segura-Cox}, {Bianchi}, {Cuello}, {L{\'o}pez-Sepulcre}, {Fedele}, {Caselli}, {Charnley}, {Johnstone}, {Zhang}, {Maureira}, {Zhang}, {Sabatini}, {Svoboda}, {Jim{\'e}nez-Serra}, {Loinard}, {Mercimek}, {Murillo}, \& {Yamamoto}}]{2024Codella}
{Codella}, C., {Podio}, L., {De Simone}, M., {et~al.} 2024, \mnras, 528, 7383, \dodoi{10.1093/mnras/stae472}

\bibitem[{{Cotton}(2017)}]{2017Cotton}
{Cotton}, W.~D. 2017, \pasp, 129, 094501, \dodoi{10.1088/1538-3873/aa793f}

\bibitem[{{Cuello} {et~al.}(2023){Cuello}, {M{\'e}nard}, \& {Price}}]{2023Cuello}
{Cuello}, N., {M{\'e}nard}, F., \& {Price}, D.~J. 2023, European Physical Journal Plus, 138, 11, \dodoi{10.1140/epjp/s13360-022-03602-w}

\bibitem[{{Davis}(1951)}]{1951Davis}
{Davis}, L. 1951, Physical Review, 81, 890, \dodoi{10.1103/PhysRev.81.890.2}

\bibitem[{Dullemond(2012)}]{2012Dullemond}
Dullemond, C.~P. 2012, Astrophysics Source Code Library, 1202.015.
\newblock \url{http://adsabs.harvard.edu/abs/2012ascl.soft02015D}

\bibitem[{{Dullemond} {et~al.}(2019){Dullemond}, {K{\"u}ffmeier}, {Goicovic}, {Fukagawa}, {Oehl}, \& {Kramer}}]{2019Dullemond}
{Dullemond}, C.~P., {K{\"u}ffmeier}, M., {Goicovic}, F., {et~al.} 2019, \aap, 628, A20, \dodoi{10.1051/0004-6361/201832632}

\bibitem[{{Dunham} {et~al.}(2014){Dunham}, {Arce}, {Mardones}, {Lee}, {Matthews}, {Stutz}, \& {Williams}}]{2014Dunham}
{Dunham}, M.~M., {Arce}, H.~G., {Mardones}, D., {et~al.} 2014, \apj, 783, 29, \dodoi{10.1088/0004-637X/783/1/29}

\bibitem[{{Dunham} {et~al.}(2008){Dunham}, {Crapsi}, {Evans}, {Bourke}, {Huard}, {Myers}, \& {Kauffmann}}]{2008Dunham}
{Dunham}, M.~M., {Crapsi}, A., {Evans}, Neal~J., I., {et~al.} 2008, \apjs, 179, 249, \dodoi{10.1086/591085}

\bibitem[{{Flores} {et~al.}(2023){Flores}, {Ohashi}, {Tobin}, {J{\o}rgensen}, {Takakuwa}, {Li}, {Lin}, {van't Hoff}, {Plunkett}, {Yamato}, {Sai (Insa Choi)}, {Koch}, {Yen}, {Aikawa}, {Aso}, {de Gregorio-Monsalvo}, {Kido}, {Kwon}, {Lee}, {Lee}, {Looney}, {Santamar{\'\i}a-Miranda}, {Sharma}, {Thieme}, {Williams}, {Han}, {Narayanan}, \& {Lai}}]{2023Flores}
{Flores}, C., {Ohashi}, N., {Tobin}, J.~J., {et~al.} 2023, \apj, 958, 98, \dodoi{10.3847/1538-4357/acf7c1}

\bibitem[{{Frerking} {et~al.}(1982){Frerking}, {Langer}, \& {Wilson}}]{1982Frerking}
{Frerking}, M.~A., {Langer}, W.~D., \& {Wilson}, R.~W. 1982, \apj, 262, 590, \dodoi{10.1086/160451}

\bibitem[{{Ginsburg} {et~al.}(2016){Ginsburg}, {Robitaille}, \& {Beaumont}}]{2016Ginsburg}
{Ginsburg}, A., {Robitaille}, T., \& {Beaumont}, C. 2016, {pvextractor: Position-Velocity Diagram Extractor}, Astrophysics Source Code Library, record ascl:1608.010.
\newblock \doeprint{1608.010}

\bibitem[{{Gong} {et~al.}(2022){Gong}, {Liu}, {Wang}, {Zhu}, {Li}, {Yang}, \& {Sun}}]{2022GOng}
{Gong}, Y., {Liu}, S., {Wang}, J., {et~al.} 2022, \aap, 663, A82, \dodoi{10.1051/0004-6361/202142713}

\bibitem[{{Goodman} {et~al.}(1993){Goodman}, {Benson}, {Fuller}, \& {Myers}}]{1993Goodman}
{Goodman}, A.~A., {Benson}, P.~J., {Fuller}, G.~A., \& {Myers}, P.~C. 1993, \apj, 406, 528, \dodoi{10.1086/172465}

\bibitem[{{Gupta} {et~al.}(2023){Gupta}, {Miotello}, {Manara}, {Williams}, {Facchini}, {Beccari}, {Birnstiel}, {Ginski}, {Hacar}, {K{\"u}ffmeier}, {Testi}, {Tychoniec}, \& {Yen}}]{2023Gupta}
{Gupta}, A., {Miotello}, A., {Manara}, C.~F., {et~al.} 2023, \aap, 670, L8, \dodoi{10.1051/0004-6361/202245254}

\bibitem[{{Han} {et~al.}(2023){Han}, {Kwon}, {Aso}, {Bae}, \& {Sheehan}}]{2023Han}
{Han}, I., {Kwon}, W., {Aso}, Y., {Bae}, J., \& {Sheehan}, P. 2023, \apj, 956, 9, \dodoi{10.3847/1538-4357/acf853}

\bibitem[{{Hanawa} {et~al.}(2024){Hanawa}, {Garufi}, {Podio}, {Codella}, \& {Segura-Cox}}]{2024Hanawa}
{Hanawa}, T., {Garufi}, A., {Podio}, L., {Codella}, C., \& {Segura-Cox}, D. 2024, \mnras, 528, 6581, \dodoi{10.1093/mnras/stae338}

\bibitem[{{Hatchell} {et~al.}(2012){Hatchell}, {Terebey}, {Huard}, {Mamajek}, {Allen}, {Bourke}, {Dunham}, {Gutermuth}, {Harvey}, {J{\o}rgensen}, {Mer{\'\i}n}, {Noriega-Crespo}, \& {Peterson}}]{2012Hatchell}
{Hatchell}, J., {Terebey}, S., {Huard}, T., {et~al.} 2012, \apj, 754, 104, \dodoi{10.1088/0004-637X/754/2/104}

\bibitem[{{Heitsch} {et~al.}(2001){Heitsch}, {Zweibel}, {Mac Low}, {Li}, \& {Norman}}]{2001Heitsch}
{Heitsch}, F., {Zweibel}, E.~G., {Mac Low}, M.-M., {Li}, P., \& {Norman}, M.~L. 2001, \apj, 561, 800, \dodoi{10.1086/323489}

\bibitem[{{Hsieh} {et~al.}(2023){Hsieh}, {Segura-Cox}, {Pineda}, {Caselli}, {Bouscasse}, {Neri}, {Lopez-Sepulcre}, {Valdivia-Mena}, {Maureira}, {Henning}, {Smirnov-Pinchukov}, {Semenov}, {M{\"o}ller}, {Cunningham}, {Fuente}, {Marino}, {Dutrey}, {Tafalla}, {Chapillon}, {Ceccarelli}, \& {Zhao}}]{2023Hsieh}
{Hsieh}, T.~H., {Segura-Cox}, D.~M., {Pineda}, J.~E., {et~al.} 2023, \aap, 669, A137, \dodoi{10.1051/0004-6361/202244183}

\bibitem[{{Huang} {et~al.}(2021){Huang}, {Bergin}, {{\"O}berg}, {Andrews}, {Teague}, {Law}, {Kalas}, {Aikawa}, {Bae}, {Bergner}, {Booth}, {Bosman}, {Calahan}, {Cataldi}, {Cleeves}, {Czekala}, {Ilee}, {Le Gal}, {Guzm{\'a}n}, {Long}, {Loomis}, {M{\'e}nard}, {Nomura}, {Qi}, {Schwarz}, {Tsukagoshi}, {van't Hoff}, {Walsh}, {Wilner}, {Yamato}, \& {Zhang}}]{2021Huang}
{Huang}, J., {Bergin}, E.~A., {{\"O}berg}, K.~I., {et~al.} 2021, \apjs, 257, 19, \dodoi{10.3847/1538-4365/ac143e}

\bibitem[{{Huard} {et~al.}(1999){Huard}, {Sandell}, \& {Weintraub}}]{1999Huard}
{Huard}, T.~L., {Sandell}, G., \& {Weintraub}, D.~A. 1999, \apj, 526, 833, \dodoi{10.1086/308022}

\bibitem[{Hunter(2007)}]{Hunter2007}
Hunter, J.~D. 2007, Computing in Science and Engineering, 9, 90, \dodoi{10.1109/MCSE.2007.55}

\bibitem[{{Imai} {et~al.}(2022){Imai}, {Oya}, {Svoboda}, {Liu}, {Lefloch}, {Viti}, {Zhang}, {Ceccarelli}, {Codella}, {Chandler}, {Sakai}, {Aikawa}, {Alves}, {Balucani}, {Bianchi}, {Bouvier}, {Busquet}, {Caselli}, {Caux}, {Charnley}, {Choudhury}, {Cuello}, {Simone}, {Dulieu}, {Dur{\'a}n}, {Evans}, {Favre}, {Fedele}, {Feng}, {Fontani}, {Francis}, {Hama}, {Hanawa}, {Herbst}, {Hirano}, {Hirota}, {Isella}, {J{\'\i}menez-Serra}, {Johnstone}, {Kahane}, {Le Gal}, {Loinard}, {L{\'o}pez-Sepulcre}, {Maud}, {Maureira}, {Menard}, {Mercimek}, {Miotello}, {Moellenbrock}, {Mori}, {Murillo}, {Nakatani}, {Nomura}, {Oba}, {O'Donoghue}, {Ohashi}, {Okoda}, {Ospina-Zamudio}, {Pineda}, {Podio}, {Rimola}, {Sakai}, {Segura-Cox}, {Shirley}, {Taquet}, {Testi}, {Vastel}, {Watanabe}, {Watanabe}, {Witzel}, {Xue}, {Zhao}, \& {Yamamoto}}]{2022Imai}
{Imai}, M., {Oya}, Y., {Svoboda}, B., {et~al.} 2022, \apj, 934, 70, \dodoi{10.3847/1538-4357/ac77e7}

\bibitem[{{Kauffmann} {et~al.}(2008){Kauffmann}, {Bertoldi}, {Bourke}, {Evans}, \& {Lee}}]{2008Kauffmann}
{Kauffmann}, J., {Bertoldi}, F., {Bourke}, T.~L., {Evans}, N.~J., I., \& {Lee}, C.~W. 2008, \aap, 487, 993, \dodoi{10.1051/0004-6361:200809481}

\bibitem[{{Kido} {et~al.}(2023){Kido}, {Takakuwa}, {Saigo}, {Ohashi}, {Tobin}, {J{\o}rgensen}, {Aikawa}, {Aso}, {Encalada}, {Flores}, {Gavino}, {de Gregorio-Monsalvo}, {Han}, {Hirano}, {Koch}, {Kwon}, {Lai}, {Lee}, {Lee}, {Li}, {Lin}, {Looney}, {Mori}, {Narayanan}, {Plunkett}, {Phuong}, {(Insa Choi)}, {Santamar{\'\i}a-Miranda}, {Sharma}, {Sheehan}, {Thieme}, {Tomida}, {van't Hoff}, {Williams}, {Yamato}, \& {Yen}}]{2023Kido}
{Kido}, M., {Takakuwa}, S., {Saigo}, K., {et~al.} 2023, \apj, 953, 190, \dodoi{10.3847/1538-4357/acdd7a}

\bibitem[{{Kuffmeier} {et~al.}(2017){Kuffmeier}, {Haugb{\o}lle}, \& {Nordlund}}]{2017Kuffmeier}
{Kuffmeier}, M., {Haugb{\o}lle}, T., \& {Nordlund}, {\r{A}}. 2017, \apj, 846, 7, \dodoi{10.3847/1538-4357/aa7c64}

\bibitem[{{Kuffmeier} {et~al.}(2023){Kuffmeier}, {Jensen}, \& {Haugb{\o}lle}}]{2023Kuffmeier}
{Kuffmeier}, M., {Jensen}, S.~S., \& {Haugb{\o}lle}, T. 2023, European Physical Journal Plus, 138, 272, \dodoi{10.1140/epjp/s13360-023-03880-y}

\bibitem[{{Lam} {et~al.}(2019){Lam}, {Li}, {Chen}, {Tomida}, \& {Zhao}}]{2019Lam}
{Lam}, K.~H., {Li}, Z.-Y., {Chen}, C.-Y., {Tomida}, K., \& {Zhao}, B. 2019, \mnras, 489, 5326, \dodoi{10.1093/mnras/stz2436}

\bibitem[{{Larson}(1969)}]{1969Larson}
{Larson}, R.~B. 1969, \mnras, 145, 271, \dodoi{10.1093/mnras/145.3.271}

\bibitem[{{Launhardt} {et~al.}(2010){Launhardt}, {Nutter}, {Ward-Thompson}, {Bourke}, {Henning}, {Khanzadyan}, {Schmalzl}, {Wolf}, \& {Zylka}}]{2010Launhardt}
{Launhardt}, R., {Nutter}, D., {Ward-Thompson}, D., {et~al.} 2010, \apjs, 188, 139, \dodoi{10.1088/0067-0049/188/1/139}

\bibitem[{{Launhardt} {et~al.}(2013){Launhardt}, {Stutz}, {Schmiedeke}, {Henning}, {Krause}, {Balog}, {Beuther}, {Birkmann}, {Hennemann}, {Kainulainen}, {Khanzadyan}, {Linz}, {Lippok}, {Nielbock}, {Pitann}, {Ragan}, {Risacher}, {Schmalzl}, {Shirley}, {Stecklum}, {Steinacker}, \& {Tackenberg}}]{2013Launhardt}
{Launhardt}, R., {Stutz}, A.~M., {Schmiedeke}, A., {et~al.} 2013, \aap, 551, A98, \dodoi{10.1051/0004-6361/201220477}

\bibitem[{{Lebreuilly} {et~al.}(2024){Lebreuilly}, {Hennebelle}, {Colman}, {Maury}, {Tung}, {Testi}, {Klessen}, {Molinari}, {Commer{\c{c}}on}, {Gonz{\'a}lez}, {Pacetti}, {Somigliana}, \& {Rosotti}}]{2024Lebreuilly}
{Lebreuilly}, U., {Hennebelle}, P., {Colman}, T., {et~al.} 2024, \aap, 682, A30, \dodoi{10.1051/0004-6361/202346558}

\bibitem[{{Lee} {et~al.}(2023){Lee}, {Matsumoto}, {Kim}, {Lee}, {Harsono}, {Bae}, {Evans}, {Inutsuka}, {Choi}, {Tatematsu}, {Lee}, \& {Jaffe}}]{2023Lee}
{Lee}, J.-E., {Matsumoto}, T., {Kim}, H.-J., {et~al.} 2023, \apj, 953, 82, \dodoi{10.3847/1538-4357/acdd5b}

\bibitem[{{Lee} {et~al.}(2024){Lee}, {Kim}, {Lee}, {Lee}, {Baek}, {Yun}, {Aikawa}, {Johnstone}, {Herczeg}, \& {Cieza}}]{2024Lee}
{Lee}, J.-E., {Kim}, C.-H., {Lee}, S., {et~al.} 2024, \apj, 966, 119, \dodoi{10.3847/1538-4357/ad3106}

\bibitem[{{Lesur} {et~al.}(2015){Lesur}, {Hennebelle}, \& {Fromang}}]{2015Lesur}
{Lesur}, G., {Hennebelle}, P., \& {Fromang}, S. 2015, \aap, 582, L9, \dodoi{10.1051/0004-6361/201526734}

\bibitem[{{Lin} {et~al.}(2024){Lin}, {Yen}, \& {Lai}}]{2024Lin}
{Lin}, S.-J., {Yen}, H.-W., \& {Lai}, S.-P. 2024, \aj, 168, 107, \dodoi{10.3847/1538-3881/ad5add}

\bibitem[{{L{\'o}pez-V{\'a}zquez} {et~al.}(2024){L{\'o}pez-V{\'a}zquez}, {Lee}, {Shang}, {Cabrit}, {Krasnopolsky}, {Codella}, {Liu}, {Podio}, {Dutta}, {Murphy}, \& {Wiseman}}]{2024Lopez-Vazquez}
{L{\'o}pez-V{\'a}zquez}, J.~A., {Lee}, C.-F., {Shang}, H., {et~al.} 2024, arXiv e-prints, arXiv:2411.01728, \dodoi{10.48550/arXiv.2411.01728}

\bibitem[{McMullin {et~al.}(2007)McMullin, Waters, Schiebel, Young, \& Golap}]{Mcmullin2007}
McMullin, J.~P., Waters, B., Schiebel, D., Young, W., \& Golap, K. 2007, Astronomical Data Analysis Software and Systems XVI, 376, 127

\bibitem[{{Mercimek} {et~al.}(2022){Mercimek}, {Codella}, {Podio}, {Bianchi}, {Chahine}, {Bouvier}, {L{\'o}pez-Sepulcre}, {Neri}, \& {Ceccarelli}}]{2022Mercimek}
{Mercimek}, S., {Codella}, C., {Podio}, L., {et~al.} 2022, \aap, 659, A67, \dodoi{10.1051/0004-6361/202141790}

\bibitem[{{Miura} {et~al.}(2017){Miura}, {Yamamoto}, {Nomura}, {Nakamoto}, {Tanaka}, {Tanaka}, \& {Nagasawa}}]{2017Miura}
{Miura}, H., {Yamamoto}, T., {Nomura}, H., {et~al.} 2017, \apj, 839, 47, \dodoi{10.3847/1538-4357/aa67df}

\bibitem[{{Momose} {et~al.}(1998){Momose}, {Ohashi}, {Kawabe}, {Nakano}, \& {Hayashi}}]{1998Momose}
{Momose}, M., {Ohashi}, N., {Kawabe}, R., {Nakano}, T., \& {Hayashi}, M. 1998, \apj, 504, 314, \dodoi{10.1086/306061}

\bibitem[{{Murillo} {et~al.}(2022){Murillo}, {van Dishoeck}, {Hacar}, {Harsono}, \& {J{\o}rgensen}}]{2022Murillo}
{Murillo}, N.~M., {van Dishoeck}, E.~F., {Hacar}, A., {Harsono}, D., \& {J{\o}rgensen}, J.~K. 2022, \aap, 658, A53, \dodoi{10.1051/0004-6361/202141250}

\bibitem[{{Nakamura} {et~al.}(2024){Nakamura}, {Nguyen-Luong}, {Ishihara}, \& {Yoshino}}]{2024Nakamura}
{Nakamura}, F., {Nguyen-Luong}, Q., {Ishihara}, K., \& {Yoshino}, A. 2024, arXiv e-prints, arXiv:2409.02661, \dodoi{10.48550/arXiv.2409.02661}

\bibitem[{{Nakano} \& {Nakamura}(1978)}]{1978Nakano}
{Nakano}, T., \& {Nakamura}, T. 1978, \pasj, 30, 671

\bibitem[{{Ohashi} {et~al.}(1997){Ohashi}, {Hayashi}, {Ho}, {Momose}, {Tamura}, {Hirano}, \& {Sargent}}]{1997Ohashi}
{Ohashi}, N., {Hayashi}, M., {Ho}, P. T.~P., {et~al.} 1997, \apj, 488, 317, \dodoi{10.1086/304685}

\bibitem[{{Ohashi} {et~al.}(2023){Ohashi}, {Tobin}, {J$\o$rgensen}, \& {eDisk~Team}}]{2023Ohashi}
{Ohashi}, N., {Tobin}, J.~J., {J$\o$rgensen}, J.~J., \& {eDisk~Team}. 2023, \apj, in~press

\bibitem[{{Ohashi} {et~al.}(2014){Ohashi}, {Saigo}, {Aso}, {Aikawa}, {Koyamatsu}, {Machida}, {Saito}, {Takahashi}, {Takakuwa}, {Tomida}, {Tomisaka}, \& {Yen}}]{2014Ohashi}
{Ohashi}, N., {Saigo}, K., {Aso}, Y., {et~al.} 2014, \apj, 796, 131, \dodoi{10.1088/0004-637X/796/2/131}

\bibitem[{{Okoda} {et~al.}(2022){Okoda}, {Oya}, {Imai}, {Sakai}, {Watanabe}, {L{\'o}pez-Sepulcre}, {Saigo}, \& {Yamamoto}}]{2022Okoda}
{Okoda}, Y., {Oya}, Y., {Imai}, M., {et~al.} 2022, \apj, 935, 136, \dodoi{10.3847/1538-4357/ac7ff4}

\bibitem[{{Ostriker} {et~al.}(2001){Ostriker}, {Stone}, \& {Gammie}}]{2001Ostriker}
{Ostriker}, E.~C., {Stone}, J.~M., \& {Gammie}, C.~F. 2001, \apj, 546, 980, \dodoi{10.1086/318290}

\bibitem[{{Oya} {et~al.}(2016){Oya}, {Sakai}, {L{\'o}pez-Sepulcre}, {Watanabe}, {Ceccarelli}, {Lefloch}, {Favre}, \& {Yamamoto}}]{2016Oya}
{Oya}, Y., {Sakai}, N., {L{\'o}pez-Sepulcre}, A., {et~al.} 2016, \apj, 824, 88, \dodoi{10.3847/0004-637X/824/2/88}

\bibitem[{{Pineda} {et~al.}(2021){Pineda}, {Schmiedeke}, {Caselli}, {Stahler}, {Frayer}, {Church}, \& {Harris}}]{2021Pineda}
{Pineda}, J.~E., {Schmiedeke}, A., {Caselli}, P., {et~al.} 2021, \apj, 912, 7, \dodoi{10.3847/1538-4357/abebdd}

\bibitem[{{Pineda} {et~al.}(2020){Pineda}, {Segura-Cox}, {Caselli}, {Cunningham}, {Zhao}, {Schmiedeke}, {Maureira}, \& {Neri}}]{2020Pineda}
{Pineda}, J.~E., {Segura-Cox}, D., {Caselli}, P., {et~al.} 2020, Nature Astronomy, 4, 1158, \dodoi{10.1038/s41550-020-1150-z}

\bibitem[{{Pineda} {et~al.}(2023){Pineda}, {Arzoumanian}, {Andre}, {Friesen}, {Zavagno}, {Clarke}, {Inoue}, {Chen}, {Lee}, {Soler}, \& {Kuffmeier}}]{2023Pineda}
{Pineda}, J.~E., {Arzoumanian}, D., {Andre}, P., {et~al.} 2023, in Astronomical Society of the Pacific Conference Series, Vol. 534, Protostars and Planets VII, ed. S.~{Inutsuka}, Y.~{Aikawa}, T.~{Muto}, K.~{Tomida}, \& M.~{Tamura}, 233, \dodoi{10.48550/arXiv.2205.03935}

\bibitem[{{Plunkett} {et~al.}(2023){Plunkett}, {Hacar}, {Moser-Fischer}, {Petry}, {Teuben}, {Pingel}, {Kunneriath}, {Takagi}, {Miyamoto}, {Moravec}, {Suri}, {Hess}, {Hoffman}, \& {Mason}}]{2023Plunkett}
{Plunkett}, A., {Hacar}, A., {Moser-Fischer}, L., {et~al.} 2023, \pasp, 135, 034501, \dodoi{10.1088/1538-3873/acb9bd}

\bibitem[{{Sai} {et~al.}(2020){Sai}, {Ohashi}, {Saigo}, {Matsumoto}, {Aso}, {Takakuwa}, {Aikawa}, {Kurose}, {Yen}, {Tomisaka}, {Tomida}, \& {Machida}}]{2020Sai}
{Sai}, J., {Ohashi}, N., {Saigo}, K., {et~al.} 2020, \apj, 893, 51, \dodoi{10.3847/1538-4357/ab8065}

\bibitem[{{Sakai} {et~al.}(2014){Sakai}, {Oya}, {Sakai}, {Watanabe}, {Hirota}, {Ceccarelli}, {Kahane}, {Lopez-Sepulcre}, {Lefloch}, {Vastel}, {Bottinelli}, {Caux}, {Coutens}, {Aikawa}, {Takakuwa}, {Ohashi}, {Yen}, \& {Yamamoto}}]{2014Sakaib}
{Sakai}, N., {Oya}, Y., {Sakai}, T., {et~al.} 2014, \apjl, 791, L38, \dodoi{10.1088/2041-8205/791/2/L38}

\bibitem[{{Seifried} {et~al.}(2015){Seifried}, {Banerjee}, {Pudritz}, \& {Klessen}}]{2015Seifried}
{Seifried}, D., {Banerjee}, R., {Pudritz}, R.~E., \& {Klessen}, R.~S. 2015, \mnras, 446, 2776, \dodoi{10.1093/mnras/stu2282}

\bibitem[{{Semenov} {et~al.}(2003){Semenov}, {Henning}, {Helling}, {Ilgner}, \& {Sedlmayr}}]{2003Semenov}
{Semenov}, D., {Henning}, T., {Helling}, C., {Ilgner}, M., \& {Sedlmayr}, E. 2003, \aap, 410, 611, \dodoi{10.1051/0004-6361:20031279}

\bibitem[{{Shu}(1977)}]{1977Shu}
{Shu}, F.~H. 1977, \apj, 214, 488, \dodoi{10.1086/155274}

\bibitem[{{Skalidis} {et~al.}(2021){Skalidis}, {Sternberg}, {Beattie}, {Pavlidou}, \& {Tassis}}]{2021Skalidis}
{Skalidis}, R., {Sternberg}, J., {Beattie}, J.~R., {Pavlidou}, V., \& {Tassis}, K. 2021, \aap, 656, A118, \dodoi{10.1051/0004-6361/202142045}

\bibitem[{{Smallwood} {et~al.}(2024){Smallwood}, {Nealon}, {Cuello}, {Dong}, \& {Booth}}]{2024Smallwood}
{Smallwood}, J.~L., {Nealon}, R., {Cuello}, N., {Dong}, R., \& {Booth}, R.~A. 2024, \mnras, 527, 2094, \dodoi{10.1093/mnras/stad3057}

\bibitem[{{Smallwood} {et~al.}(2023){Smallwood}, {Yang}, {Zhu}, {Martin}, {Dong}, {Cuello}, \& {Isella}}]{2023Smallwood}
{Smallwood}, J.~L., {Yang}, C.-C., {Zhu}, Z., {et~al.} 2023, \mnras, 521, 3500, \dodoi{10.1093/mnras/stad742}

\bibitem[{{Stahler}(1988)}]{1988Stahler}
{Stahler}, S.~W. 1988, \apj, 332, 804, \dodoi{10.1086/166694}

\bibitem[{{Tachihara} {et~al.}(2000){Tachihara}, {Mizuno}, \& {Fukui}}]{2000Tachihara}
{Tachihara}, K., {Mizuno}, A., \& {Fukui}, Y. 2000, \apj, 528, 817, \dodoi{10.1086/308189}

\bibitem[{{Takakuwa} {et~al.}(2013){Takakuwa}, {Saito}, {Lim}, \& {Saigo}}]{2013Takakuwa}
{Takakuwa}, S., {Saito}, M., {Lim}, J., \& {Saigo}, K. 2013, \apj, 776, 51, \dodoi{10.1088/0004-637X/776/1/51}

\bibitem[{{Takakuwa} {et~al.}(2018){Takakuwa}, {Tsukamoto}, {Saigo}, \& {Saito}}]{2018Takakuwa}
{Takakuwa}, S., {Tsukamoto}, Y., {Saigo}, K., \& {Saito}, M. 2018, \apj, 865, 51, \dodoi{10.3847/1538-4357/aadb93}

\bibitem[{{Takakuwa} {et~al.}(2024){Takakuwa}, {Saigo}, {Kido}, {Ohashi}, {Tobin}, {J{\o}rgensen}, {Aikawa}, {Aso}, {Gavino}, {Han}, {Koch}, {Kwon}, {Lee}, {Lee}, {Li}, {Lin}, {Looney}, {Mori}, {Sai}, {Sharma}, {Sheehan}, {Tomida}, {Williams}, {Yamato}, \& {Yen}}]{2024Takakuwa}
{Takakuwa}, S., {Saigo}, K., {Kido}, M., {et~al.} 2024, \apj, 964, 24, \dodoi{10.3847/1538-4357/ad1f57}

\bibitem[{{Taniguchi} {et~al.}(2024){Taniguchi}, {Pineda}, {Caselli}, {Shimoikura}, {Friesen}, {Segura-Cox}, \& {Schmiedeke}}]{2024Taniguchi}
{Taniguchi}, K., {Pineda}, J.~E., {Caselli}, P., {et~al.} 2024, \apj, 965, 162, \dodoi{10.3847/1538-4357/ad2fa1}

\bibitem[{{Terebey} {et~al.}(1984){Terebey}, {Shu}, \& {Cassen}}]{1984Terebey}
{Terebey}, S., {Shu}, F.~H., \& {Cassen}, P. 1984, \apj, 286, 529, \dodoi{10.1086/162628}

\bibitem[{{Thieme} {et~al.}(2022){Thieme}, {Lai}, {Lin}, {Cheong}, {Lee}, {Yen}, {Li}, {Lam}, \& {Zhao}}]{2022Thieme}
{Thieme}, T.~J., {Lai}, S.-P., {Lin}, S.-J., {et~al.} 2022, \apj, 925, 32, \dodoi{10.3847/1538-4357/ac382b}

\bibitem[{{Tobin} {et~al.}(2012){Tobin}, {Hartmann}, {Chiang}, {Wilner}, {Looney}, {Loinard}, {Calvet}, \& {D'Alessio}}]{2012Tobin}
{Tobin}, J.~J., {Hartmann}, L., {Chiang}, H.-F., {et~al.} 2012, \nat, 492, 83, \dodoi{10.1038/nature11610}

\bibitem[{{Tobin} \& {Sheehan}(2024)}]{2024Tobin}
{Tobin}, J.~J., \& {Sheehan}, P.~D. 2024, arXiv e-prints, arXiv:2403.15550, \dodoi{10.48550/arXiv.2403.15550}

\bibitem[{{Tokuda} {et~al.}(2014){Tokuda}, {Onishi}, {Saigo}, {Kawamura}, {Fukui}, {Matsumoto}, {Inutsuka}, {Machida}, {Tomida}, \& {Tachihara}}]{2014Tokuda}
{Tokuda}, K., {Onishi}, T., {Saigo}, K., {et~al.} 2014, \apjl, 789, L4, \dodoi{10.1088/2041-8205/789/1/L4}

\bibitem[{{Tu} {et~al.}(2023){Tu}, {Li}, {Lam}, {Tomida}, \& {Hsu}}]{2023Tu}
{Tu}, Y., {Li}, Z.-Y., {Lam}, K.~H., {Tomida}, K., \& {Hsu}, C.-Y. 2023, arXiv e-prints, arXiv:2307.16774, \dodoi{10.48550/arXiv.2307.16774}

\bibitem[{{Ulrich}(1976)}]{1976Ulrich}
{Ulrich}, R.~K. 1976, \apj, 210, 377, \dodoi{10.1086/154840}

\bibitem[{{Valdivia-Mena} {et~al.}(2022){Valdivia-Mena}, {Pineda}, {Segura-Cox}, {Caselli}, {Neri}, {L{\'o}pez-Sepulcre}, {Cunningham}, {Bouscasse}, {Semenov}, {Henning}, {Pi{\'e}tu}, {Chapillon}, {Dutrey}, {Fuente}, {Guilloteau}, {Hsieh}, {Jim{\'e}nez-Serra}, {Marino}, {Maureira}, {Smirnov-Pinchukov}, {Tafalla}, \& {Zhao}}]{2022Valdivia}
{Valdivia-Mena}, M.~T., {Pineda}, J.~E., {Segura-Cox}, D.~M., {et~al.} 2022, \aap, 667, A12, \dodoi{10.1051/0004-6361/202243310}

\bibitem[{{Vall{\'e}e} {et~al.}(2000){Vall{\'e}e}, {Bastien}, \& {Greaves}}]{2000Vallee}
{Vall{\'e}e}, J.~P., {Bastien}, P., \& {Greaves}, J.~S. 2000, \apj, 542, 352, \dodoi{10.1086/309531}

\bibitem[{{Vall{\'e}e} \& {Fiege}(2007)}]{2007Vallee}
{Vall{\'e}e}, J.~P., \& {Fiege}, J.~D. 2007, \aj, 134, 628, \dodoi{10.1086/519163}

\bibitem[{{Vall{\'e}e} {et~al.}(2003){Vall{\'e}e}, {Greaves}, \& {Fiege}}]{2003Vallee}
{Vall{\'e}e}, J.~P., {Greaves}, J.~S., \& {Fiege}, J.~D. 2003, \apj, 588, 910, \dodoi{10.1086/374309}

\bibitem[{{van Gelder} {et~al.}(2021){van Gelder}, {Tabone}, {van Dishoeck}, \& {Godard}}]{2021vanGelder}
{van Gelder}, M.~L., {Tabone}, B., {van Dishoeck}, E.~F., \& {Godard}, B. 2021, \aap, 653, A159, \dodoi{10.1051/0004-6361/202141591}

\bibitem[{{van't Hoff} {et~al.}(2023){van't Hoff}, {Tobin}, {Li}, {Ohashi}, {J{\o}rgensen}, {Lin}, {Aikawa}, {Aso}, {de Gregorio-Monsalvo}, {Gavino}, {Han}, {Koch}, {Kwon}, {Lee}, {Lee}, {Looney}, {Narayanan}, {Plunkett}, {Sai}, {Santamar{\'\i}a-Miranda}, {Sharma}, {Sheehan}, {Takakuwa}, {Thieme}, {Williams}, {Lai}, {Phuong}, \& {Yen}}]{2023van'tHoff}
{van't Hoff}, M. L.~R., {Tobin}, J.~J., {Li}, Z.-Y., {et~al.} 2023, \apj, 951, 10, \dodoi{10.3847/1538-4357/accf87}

\bibitem[{{Wang} {et~al.}(1995){Wang}, {Evans}, {Zhou}, \& {Clemens}}]{1995Wang}
{Wang}, Y., {Evans}, Neal~J., I., {Zhou}, S., \& {Clemens}, D.~P. 1995, \apj, 454, 217, \dodoi{10.1086/176478}

\bibitem[{{Wu} {et~al.}(1996){Wu}, {Huang}, \& {He}}]{1996Wu}
{Wu}, Y., {Huang}, M., \& {He}, J. 1996, \aaps, 115, 283

\bibitem[{{Yamato} {et~al.}(2023){Yamato}, {Aikawa}, {Ohashi}, {Tobin}, {J{\o}rgensen}, {Takakuwa}, {Aso}, {Sai}, {Flores}, {de Gregorio-Monsalvo}, {Hirano}, {Han}, {Kido}, {Koch}, {Kwon}, {Lai}, {Lee}, {Lee}, {Li}, {Lin}, {Looney}, {Mori}, {Narayanan}, {Phuong}, {Saigo}, {Santamar{\'\i}a-Miranda}, {Sharma}, {Thieme}, {Tomida}, {van't Hoff}, \& {Yen}}]{2023Yamato}
{Yamato}, Y., {Aikawa}, Y., {Ohashi}, N., {et~al.} 2023, \apj, 951, 11, \dodoi{10.3847/1538-4357/accd71}

\bibitem[{{Yano} {et~al.}(2024){Yano}, {Nakamura}, \& {Kinoshita}}]{2024Yano}
{Yano}, Y., {Nakamura}, F., \& {Kinoshita}, S.~W. 2024, \apj, 964, 119, \dodoi{10.3847/1538-4357/ad2a54}

\bibitem[{{Yen} {et~al.}(2019){Yen}, {Gu}, {Hirano}, {Koch}, {Lee}, {Liu}, \& {Takakuwa}}]{2019Yen}
{Yen}, H.-W., {Gu}, P.-G., {Hirano}, N., {et~al.} 2019, \apj, 880, 69, \dodoi{10.3847/1538-4357/ab29f8}

\bibitem[{{Yen} {et~al.}(2017){Yen}, {Koch}, {Takakuwa}, {Krasnopolsky}, {Ohashi}, \& {Aso}}]{2017Yen}
{Yen}, H.-W., {Koch}, P.~M., {Takakuwa}, S., {et~al.} 2017, \apj, 834, 178, \dodoi{10.3847/1538-4357/834/2/178}

\bibitem[{{Yen} {et~al.}(2014){Yen}, {Takakuwa}, {Ohashi}, {Aikawa}, {Aso}, {Koyamatsu}, {Machida}, {Saigo}, {Saito}, {Tomida}, \& {Tomisaka}}]{2014Yen}
{Yen}, H.-W., {Takakuwa}, S., {Ohashi}, N., {et~al.} 2014, \apj, 793, 1, \dodoi{10.1088/0004-637X/793/1/1}

\bibitem[{{Yen} {et~al.}(2024){Yen}, {Williams}, {Sai}, {Koch}, {Han}, {J{\o}rgensen}, {Kwon}, {Lee}, {Li}, {Looney}, {Narang}, {Ohashi}, {Takakuwa}, {Tobin}, {de Gregorio-Monsalvo}, {Lai}, {Lee}, \& {Tomida}}]{2024Yen}
{Yen}, H.-W., {Williams}, J.~P., {Sai}, J., {et~al.} 2024, \apj, 969, 125, \dodoi{10.3847/1538-4357/ad4c6b}

\bibitem[{{Zucker} {et~al.}(2020){Zucker}, {Speagle}, {Schlafly}, {Green}, {Finkbeiner}, {Goodman}, \& {Alves}}]{2020Zucker}
{Zucker}, C., {Speagle}, J.~S., {Schlafly}, E.~F., {et~al.} 2020, \aap, 633, A51, \dodoi{10.1051/0004-6361/201936145}

\end{thebibliography}

\end{document}